\documentclass[11pt]{article}

\usepackage{jheppub}
\usepackage{wasysym}

\usepackage{tikz-cd}
\usepackage{tabularx}
\usepackage{physics}
\usepackage{tikz}
\usepackage{graphicx} 
\usepackage{caption}
\usepackage{subcaption}
\usetikzlibrary{calc}
\usetikzlibrary{arrows}
\usetikzlibrary { decorations.pathmorphing, decorations.pathreplacing, decorations.shapes}

\def \ncp{\omega^{\raisebox{-0pt}{\includegraphics[scale=0.12]{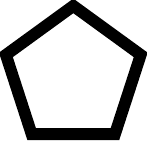}}}}
\def \cpp{\omega^{\raisebox{-0pt}{\includegraphics[scale=0.12]{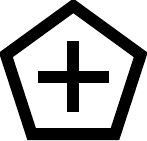}}}}

\usepackage{ae}

\usepackage{tabularx}

\usepackage{graphicx}

\usepackage{amsfonts,amsmath,amsthm,amssymb,amsbsy}

\usepackage{mathtools}

\newcommand{\arrowIn}{
\tikz \draw[-stealth] (-1pt,0) -- (1pt,0);
}

\title{The Two-loop MHV Momentum Amplituhedron from Fibrations of Fibrations}

\author[]{Livia Ferro,}\emailAdd{l.ferro@herts.ac.uk}
\author[]{Ross Glew,}\emailAdd{r.glew@herts.ac.uk}
\author[]{Tomasz \L ukowski,}\emailAdd{t.lukowski@herts.ac.uk}
\author[]{and Jonah Stalknecht}\emailAdd{j.stalknecht@herts.ac.uk}

\affiliation[]{Department of Physics, Astronomy and Mathematics, \\ University of Hertfordshire, \\  Hatfield, Hertfordshire, AL10 9AB, United Kingdom}

\abstract{Recently, a new approach to computing the canonical forms of the momentum amplituhedron in dual-momentum space was proposed by the authors. These are relevant for the integrands of scattering amplitudes in planar $\mathcal{N}=4$ super-Yang-Mills. At one-loop the idea was to view the set of all loop momenta, which we refer to as the {\it one-loop fiber} geometry, as a fibration over the tree-level kinematic data. This led to the notion of {\it tree-level chambers}, subsets of the tree-level kinematic space for which the combinatorial structure of the one-loop fiber remains unchanged, that allowed for a novel representation of the one-loop integrand. The goal of this paper is to extend these ideas to two loops for MHV integrands. Our approach will be to view the geometry accessed by the second loop momentum, similarly referred to as the {\it two-loop fiber} geometry, as a fibration over both the one-loop kinematic data and the position of the first loop momentum in the one-loop fiber. This will lead to the notion of {\it one-loop chambers}, subsets of the one-loop fibers for which the combinatorial structure of the two-loop fiber remains unchanged. We will characterise the full set of one-loop chambers and their corresponding two-loop fibers and present formulae for their canonical forms. Ultimately, this will result in a new formula for the two-loop MHV integrand written as a {\it fibration of fibration}.
}

\begin{document}

\maketitle


\section{Introduction}

\begin{figure}[h!]
\begin{center}
\begin{tikzpicture}
      \foreach \i in {1,2,3} {
	     \coordinate (V\i) at (\i*360/3+30:5);
      }
    \def \size{3};
    \def \rat{0.5};
    \draw[draw, fill=gray!10] (V1) ellipse (\size cm and \size*\rat cm) node {$\begin{matrix}\text{Spinor-helicity}\\(\lambda,\tilde{\lambda})\\ \text{Momentum amplituhedron} \end{matrix}$};
    \draw[draw, fill=gray!20] ($0.5*\rat*(V2)$) ellipse (\size cm and \size*\rat cm) node {$\begin{matrix}\text{Momentum twistors}\\(\lambda,z)\\ \text{Amplituhedron} \end{matrix}$};
    \draw[draw, fill=gray!20] (V3) ellipse (\size cm and \size*\rat cm) node {$\begin{matrix}\text{Dual momentum space}\\(\lambda,x)\\ \text{Null-cone geometries} \end{matrix}$};
	\draw[<->, thick] ($0.6*(V1)+0.4*0.5*\rat*(V2)$) --($0.4*(V1)+0.6*0.5*\rat*(V2)$);
	\draw[<->, thick] ($0.6*(V3)+0.4*0.5*\rat*(V2)$) --($0.4*(V3)+0.6*0.5*\rat*(V2)$);
	\draw[<->, thick] ($0.6*(V1)+0.4*(V3)$) --($0.4*(V1)+0.6*(V3)$);
\end{tikzpicture}
\end{center}
\caption{Triality of kinematic spaces: spinor helicity space, momentum twistor space and dual momentum space; and the corresponding positive geometries.}
\label{fig:triality}
\end{figure}
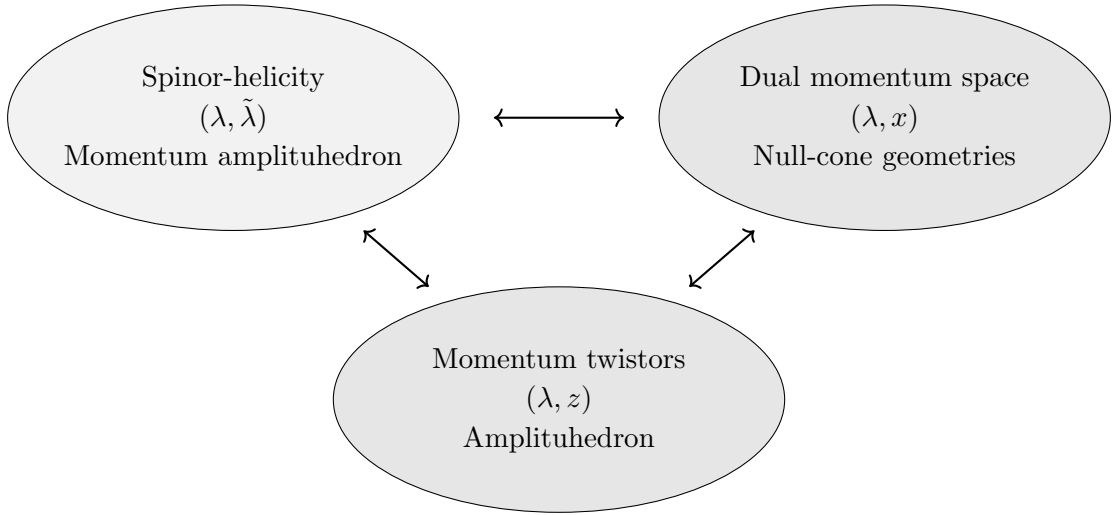

In recent years positive geometries \cite{Arkani-Hamed:2017tmz} have become an established tool in the computation of physical observables. This approach has seen most progress in the study of scattering amplitudes, where the main effort has been made for three theories in particular: planar maximally supersymmetric $\mathcal{N}=4$ super-Yang-Mills (sYM) in four dimensions \cite{Arkani-Hamed:2013jha}, ABJM in three dimensions \cite{He:2023rou} and scalar $\phi^3$ in any dimension \cite{Arkani-Hamed:2017mur}, with recent advances in more realistic theories including non-linear sigma models and pure Yang-Mills theory \cite{Arkani-Hamed:2024nhp,Arkani-Hamed:2023jry}. This paper focuses on planar $\mathcal{N}=4$ sYM for which the known positive geometries provide tree-level amplitudes and loop-level integrands. The set up of the positive geometry framework strongly depends on the kinematic space in which it is defined, and for $\mathcal{N}=4$ sYM there have been two natural kinematic spaces which have played a dominant role in recent years. One is momentum twistor space, which provides a natural description of Wilson loops in $\mathcal{N}=4$ sYM, which are known to be dual to scattering amplitudes in the planar limit. The positive geometry in momentum twistor space is the amplituhedron \cite{Arkani-Hamed:2013jha},  which provided the starting point for the development of many ideas in the field. Alternatively, one can use spinor helicity space, in which the momentum amplituhedron is defined \cite{Damgaard:2019ztj,Ferro:2022abq}. More recently, the latter has been directly translated to the  four-dimensional dual momentum space with $(2,2)$ signature. In particular, it was shown in \cite{Ferro:2023qdp} that integrands can be obtained as canonical differential forms of curvy versions of simple polytopes that are defined using the null structure of $\mathbb{R}^{2,2}$, see also \cite{Lukowski:2023nnf} for a parallel development in ABJM theory. This leads to the triality between kinematic spaces and their corresponding positive geometries illustrated in Fig.\ref{fig:triality}. 

\begin{figure}[h!]
\begin{center}
\includegraphics[scale=0.15]{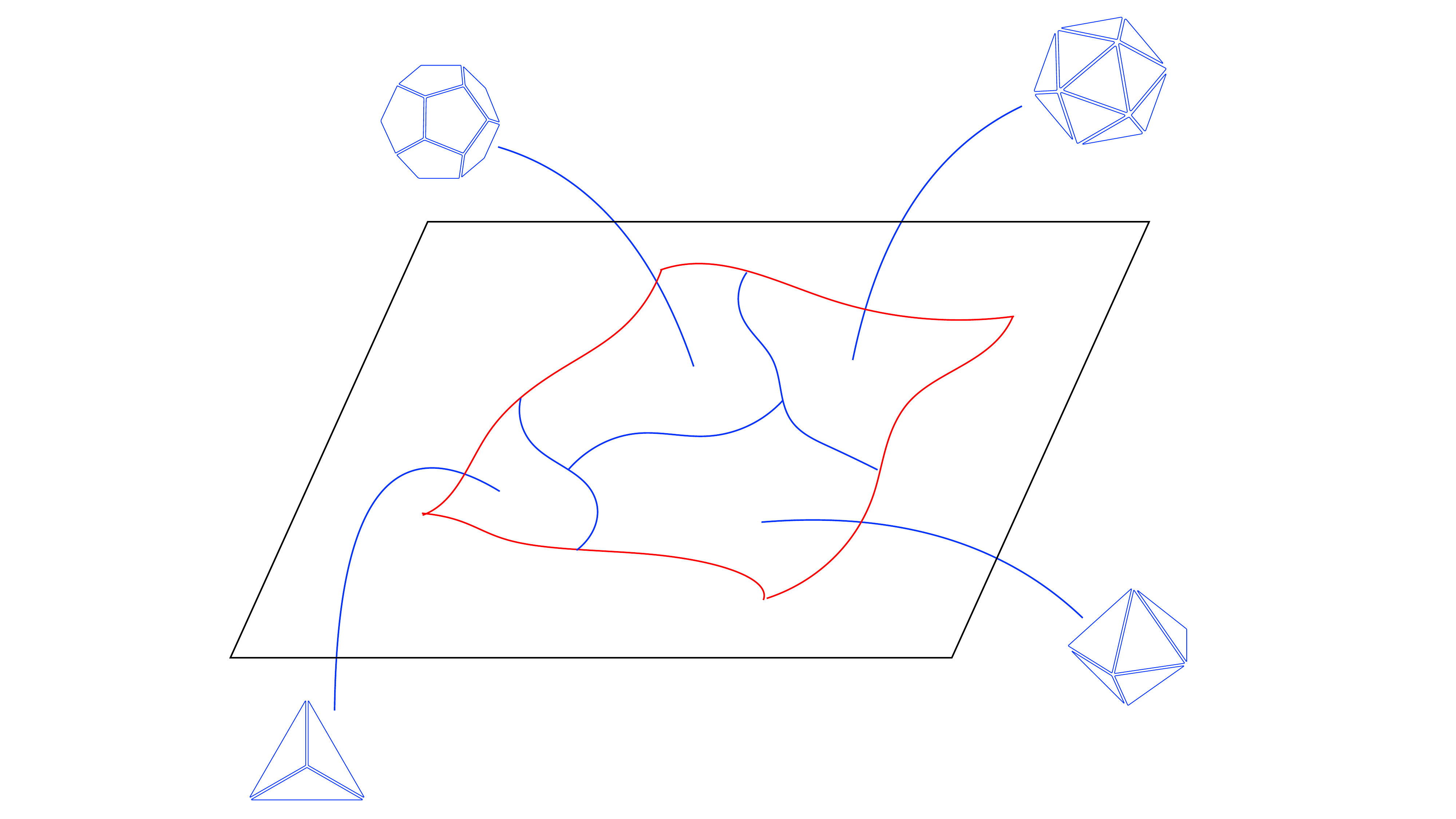}
\end{center}
\caption{One-loop geometry as a fibration over tree level. The red outlined shape indicates the tree-level geometry which is divided into tree-level chambers. Over each point in the tree-level geometry, there is a one-loop fiber geometry whose combinatorial structure is constant inside each chamber. }
\label{fig:fibration}
\end{figure}

The dual momentum space approach adopted in this paper focuses on the loop integrands of N$^{(k-2)}$MHV$_n$ amplitudes in planar $\mathcal{N}=4$ sYM. To each fixed configuration of momenta in the scattering process, one associates a null polygon, with the scattering momenta given by differences of its consecutive vertices. Then for each such polygon one defines a {\it one-loop fiber} as the set of points positively separated from each vertex of the polygon, together with a sign flip condition. For each null polygon, the so defined space turns out to be a compact region in $\mathbb{R}^{2,2}$, with vertices including the corners of the null polygon, together with a collection of quadruple intersections of null-cones. For MHV integrands, the one-loop fiber is combinatorially equivalent for all admissible null polygons. Beyond MHV, the structure of the one-loop fiber will depend upon the null polygon which is chosen, however, there exists a finite number of combinatorially inequivalent one-loop fibers. The tree-level regions for which the one-loop fibers are combinatorially equivalent were referred to as {\it chambers} in \cite{Ferro:2023qdp}, which we now refer to as {\it tree-level chambers}, and they allow for the one-loop geometry to be decomposed as a fibration over tree-level, as illustrated in Fig.~\ref{fig:fibration}.
Importantly, this allows one to write the canonical form for the one-loop integrand \cite{Bourjaily:2013mma} in a factorised way:
\begin{equation}\label{eq:intro-factor}
\Omega_{n,k,1}=\sum_{i}\Omega[\text{tree-level chamber}_i]\wedge \Omega[\text{one-loop fiber}_i],
\end{equation}
where the sum runs over all tree-level chambers for a given number of particles $n$ and helicity $k$. It is important to note that for MHV integrands, namely $k=2$, which will be the focus of this paper, there is a single tree-level chamber and hence the above sum collapses to a single term
\begin{align}
\Omega_{n,2,1}=\Omega_{n,2,0} \wedge \Omega[\text{one-loop fiber}].
\end{align}

The goal of this paper is to extend the idea of fibrations to the two-loop problem where we focus on the case of MHV integrands. From the definition provided in \cite{Ferro:2023qdp}, the two-loop geometry comprises of all pairs of points $(y_1,y_2)$ both inside the one-loop fiber, and additionally constrained to be positively separated. In contrast to \cite{He:2023rou} we will not fibrate the two-loop geometry over the tree-level. Instead, we choose to work recursively by loop order, an approach we refer to as a {\it fibration of fibration}. To achieve this we first fix the tree-level configuration given by a null polygon with vertices $x_a$, $a=1,\ldots,n$, and then for a fixed point $y_1$ inside the one-loop fiber, we study the set of points $y_2$, also inside the one-loop fiber, which are positively separated from $y_1$. We refer to this set of points as the {\it two-loop fiber} geometry with respect to $x_a$ and $y_1$. It is then a simple matter to extend the notion of chambers to loop level by defining {\it one-loop chambers} to be the set of all points $y_1$ inside the one-loop fiber for which the corresponding two-loop fibers are combinatorially equivalent. This decomposition of the one-loop fiber will allow us to produce new formulae for the one-loop MHV integrands written as a sum over one-loop chambers, and will ultimately lead to a factorised form for the two-loop MHV integrand \cite{Arkani-Hamed:2010zjl,Arkani-Hamed:2010pyv,Bourjaily:2015jna} as 
\begin{align}
\Omega_{n,2,2} =\Omega_{n,2,0} \wedge \sum_{i} \Omega\left[\text{one-loop chamber}_i\right] \wedge \Omega\left[\text{two-loop fiber}_i\right].
\label{eq:intr_fibrations}
\end{align}

Interestingly, for MHV the one-loop chambers that we find are in one-to-one correspondence to the chambers found in \cite{Parisi:2021oql} for the toy version of the amplituhedron $\mathcal{A}_{n,2,2}$. In particular, the number of one-loop chambers for the MHV$_n$ integrand are the Eulerian numbers $E_{3,n-1}$ listed in Table \ref{tab:Eulerian}. Using the results of \cite{Parisi:2021oql} we will show how the one-loop chambers are characterised by the signs of the distances of $y_1$ to each vertex of the one-loop fiber, and demonstrate how this information can be encoded simply by a permutation. We will go a step further and compute the canonical forms of the one-loop chambers and their corresponding two-loop fibers and show how they can be written as a sum over their vertices making them into curvy versions of simple polytopes. 
\begin{table}[t]
\begin{center}
\begin{tabular}{c|c|c|c|c|c}
$n$&4&5&6&7&8\\
\hline
\# one-loop chambers&1&11&66&302&1191

\end{tabular}
\end{center}
\caption{Number of one-loop chambers for MHV amplitudes.}
\label{tab:Eulerian}
\end{table}

The paper is organised as follows. In Section \ref{sec:kinematic_space} we set up the problem of studying the momentum amplituhedron in dual momentum space. In Section \ref{sec:one_loop} we move on to discuss the one-loop geometry and review the notion of tree-level chambers and one-loop fibers which are needed to understand the one-loop integrand as a fibration over tree-level. Section \ref{sec:two_loops} contains the main results of the paper. It is here we will introduce the concept of fibrations of fibrations together with the notion of one-loop chambers and two-loop fibers. We show how the canonical forms of both chambers and fibers can be written as a sum over vertices and provide concrete examples by working out the five-point case in full detail. By making use of the results of \cite{Parisi:2021oql} it will be possible to determine the full set of one-loop chambers for all MHV integrands, allowing us to extend the five-point results to arbitrary $n$. The main result will be the two-loop MHV integrand written explicitly in the fibration of fibration form \eqref{eq:fibrations}. Finally, we conclude with a discussion of our results and future research directions in Section \ref{sec:conc}.

\section{Tree-Level}
 \label{sec:kinematic_space}
 
 We start by recalling basic notions associated to the split-signature space $\mathbb{R}^{2,2}$ and explain how to describe the four-dimensional scattering kinematics relevant for planar theories.
 
\subsection{Kinematics}\label{subsec:kinematics}
To describe scattering processes in planar $\mathcal{N}=4$ sYM, we will use the framework of positive geometries and therefore consider the scattering data to be in the split-signature space $\mathbb{R}^{2,2}$, where we take the signature to be $(+,+,-,-)$. The scattering data for $n$-particle massless scattering is encoded by a set of $n$ four-dimensional momenta $p_a^\mu$, where $a=1,\ldots,n$ and $\mu=1,\ldots,4$, subject to the on-shell condition $p_a^2=0$ and momentum conservation 
\begin{equation}\label{eq:mom-cons}
\sum_{a=1}^np_a^\mu=0\,.
\end{equation} 
In the planar theory, this data can be equivalently encoded using dual momentum coordinates $x_a^\mu$ defined as 
\begin{equation}\label{eq.dual}
p_a^\mu = x_{a+1}^\mu-x_{a}^\mu\,,
\end{equation}
with the $x_a$ subject to the periodic boundary condition $x_{n+1}\equiv x_1$. Two points $x^\mu, y^\mu\in\mathbb{R}^{2,2}$ are {\it null separated} if
\begin{equation}
(x-y)^2:=(x^1 -y^1)^2+(x^2- y^2)^2-(x^3- y^3)^2-(x^4-y^4)^2=0,
\end{equation} 
otherwise they are {\it positively separated} if $(x-y)^2>0$, and they are {\it negatively separated} if $(x-y)^2<0$.  
The collection of dual momenta $x_a$ associated to a scattering process defines a null polygon in $\mathbb{R}^{2,2}$, where consecutive points $x_a$ and $x_{a+1}$ are null-separated. We denote the space of all null polygons in $\mathbb{R}^{2,2}$ with $n$ vertices by $\mathcal{P}_n$, and label each null polygon by the coordinates of its vertices $\mathbf{x}:=(x_1,x_2,\ldots,x_n)\in \mathcal{P}_n$. Note that the definition of the dual coordinates \eqref{eq.dual} is invariant under shifts of the $x_a$ by an arbitrary constant vector and for convenience we can choose $x_1=0$. This allows us to invert relation \eqref{eq.dual} to get
\begin{equation}\label{eq:p-to-dual}
x_b^\mu=\sum_{a=1}^{b-1}p_a^\mu\,.
\end{equation}

In the following we will be interested in the situation where all $x$'s are positively or null separated. A  generic set of four such dual momenta, $x_a,x_b,x_c,x_d$ for some $a,b,c,d=1,\ldots,n$, defines two points $q_{abcd}^\pm$ that are null-separated from all four of them:
\begin{equation}
\label{eq:quadcuts}
(q_{abcd}^\pm-x_a)^2=(q_{abcd}^\pm-x_b)^2=(q_{abcd}^\pm-x_c)^2=(q_{abcd}^\pm-x_d)^2=0,
\end{equation}
where the two solutions are distinguished by the sign of the determinant
\begin{equation}
\label{eq:det}
\epsilon_{abcd}(q)= \begin{tabular}{|ccccc|}
1&1&1&1&1\\
$x_a^\mu$&$x_b^\mu$&$x_c^\mu$&$x_d^\mu$&$q^\mu$
\end{tabular}\,,
\end{equation}
where $\text{sgn}(\epsilon_{abcd}(q_{abcd}^\pm))=\pm 1$. In the following, we will refer to setting a distance between two points to zero, i.e. $(y_1-y_2)^2=0$, as {\it cutting a propagator}. Then, the points $q_{abcd}^\pm$ correspond to the maximal cut solutions in four dimensions, i.e. {\it quadruple-cut points}, and will play an important role in the following as they are the vertices of the one-loop fibers defined in the next section. 

Additionally, the on-shell condition $p_a^2=0$ can be resolved by introducing spinor helicity variables and writing 
\begin{equation}\label{eq:lambda-to-p}
p^{\alpha\dot \alpha}=\begin{pmatrix}
p^0+p^2&p^1+p^3\\-p^1+p^3&p^0-p^2
\end{pmatrix}\:=\lambda^\alpha\tilde\lambda^{\dot \alpha}\,,
\end{equation}
where $\alpha=1,2$, $\dot \alpha=1,2$, and $\lambda$, $\tilde\lambda$ are real variables defined up to little group rescaling $\lambda \to t \lambda$, $\tilde\lambda\to t^{-1}\tilde\lambda$ for $t\in\mathbb{R} $. Then, each scattering process is determined by a pair of $2\times n$ matrices $(\lambda,\tilde\lambda)$ that, due to momentum conservation \eqref{eq:mom-cons}, are orthogonal to each other $\lambda \tilde\lambda^T=0$. We denote the space of all such pairs $(\lambda,\tilde\lambda)$ by $ \mathcal{K}_n$. In the following we will make use of the familiar spinor brackets
\begin{align}
\langle ab\rangle=\lambda^1_a\lambda^2_b-\lambda^2_a\lambda_b^1\,,\qquad \qquad[ ab]=\tilde\lambda^1_a\tilde\lambda^2_b-\tilde\lambda^2_a\tilde\lambda_b^1\,,
\end{align}
and  Mandelstam variables
\begin{equation}
s_{a_1,a_2,\ldots,a_r}=(p_{a_1}+p_{a_2}+\ldots p_{a_r})^2\,.
\end{equation}
In the planar theory, we are mostly interested in the case when the indices of the Mandelstam variables are consecutive. One can write these planar Mandelstam variables as distances between points in dual momentum space
\begin{equation}
s_{a,a+1,\ldots,b-1}=(x_a-x_b)^2\,.
\end{equation}

\subsection{Tree-Level Momentum Amplituhedron in Dual Momentum Space}
The relations \eqref{eq:p-to-dual} and \eqref{eq:lambda-to-p} provide a map from spinor helicity kinematic space $\mathcal{K}_n$ to the space of null polygons in dual momentum space $\mathcal{P}_n$:
\begin{equation}\label{eq:lambda-to-x}
\mathcal{K}_n \ni(\lambda,\tilde\lambda)\mapsto \mathbf{x}_{(\lambda,\tilde\lambda)}\in \mathcal{P}_n.
\end{equation} 
Following \cite{Ferro:2023qdp}, we will only be interested in a particular subset of null polygons, namely those which correspond to the images of points $(\lambda,\tilde\lambda)$ inside the tree-level momentum amplituhedron $\mathcal{M}_{n,k,0}$. The null polygons that are images of $(\lambda,\tilde\lambda)\in\mathcal{M}_{n,k,0}$, for $k=2,3,\ldots,n-2$,  through the map \eqref{eq:lambda-to-x} are relevant for the N$^{(k-2)}$MHV$_n$ amplitude, and we will denote the set of such null polygons by $\mathcal{P}_{n,k}$. An alternative characterisation of $\mathcal{P}_{n,k}$ is given by the following: for fixed $(\lambda,\tilde\lambda)\in \mathcal{K}_n$ such that
\begin{itemize}
\item all consecutive brackets of $\lambda$ are positive $\langle aa+1\rangle>0$,
\item and the sequences of brackets
$$
\{\langle a\,a+1\rangle,\langle a\,a+2\rangle,\ldots,\langle a\,a-1\rangle\},
$$
have $k-2$ sign flips for all $a=1,\ldots,n$,
\end{itemize}
a null polygon $\mathbf{x}_{(\lambda,\tilde\lambda)}$ is inside $\mathcal{P}_{n,k}$ if its vertices satisfy the following conditions:
\begin{itemize}
\item all non-consecutive vertices of $\mathbf{x}_{(\lambda,\tilde\lambda)}$ are positively separated
\begin{equation}
(x_a-x_b)^2> 0 \text{ for all } |a-b|>1,
\end{equation}
\item and the sequences of distances
$$\{\langle a+1\,a+2\rangle(x_a-\ell^*_{a+1\,a+2})^2,\langle a+1\,a+3\rangle(x_a-\ell^*_{a+1\,a+3})^2,\ldots,\langle a+1\,a-2\rangle(x_a-\ell^*_{a+1\,a-2})^2\},$$
 have $k-2$ sign flips for all $a=1,\ldots,n$, where we have defined
\begin{equation}\label{eq:lij}
\begin{cases}
\ell_{ab}^*=q_{aa+1bb+1}^+,&|a-b|>1,\\ \ell^*_{aa+1}=x_{a+1}\,.
\end{cases}
\end{equation}
We pick up a factor of $(-1)^{k-1}$ for $\langle a\, b\rangle$ when $b>n$ due to the \emph{twisted cyclic symmetry}, see \cite{Arkani-Hamed:2017vfh} for details.
\end{itemize}
The latter characterisation allows one to generate points in $\mathcal{P}_{n,k}$ without referring back to the definition of the momentum amplituhedron. 

For fixed $n$ and $k$ the set $\mathcal{P}_{n,k}$ forms an infinite family of null polygons. However, the null polygons in $\mathcal{P}_{n,k}$ can be organised into equivalence classes 
which were referred to as {\it chambers} \cite{Ferro:2023qdp}, but which we now refer to as {\it tree-level chambers}.  The tree-level chambers can be characterised by the condition that their one-loop fiber geometries, whose definition we will turn to in the next section, are combinatorially equivalent. Importantly, for MHV amplitudes, {\it all} null polygons in $\mathcal{P}_{n,2}$ belong to the same equivalence class, and therefore there exists only one tree-level chamber in this case. This means that there is a unique one-loop fiber geometry, whose combinatorial structure can be easily found for all $n$.

\section{One-Loop Review}
\label{sec:one_loop}
In this section we revisit the results of \cite{Ferro:2023qdp} for the one-loop integrand. We begin by defining the one-loop fiber, and list the full set of quadruple-cut points which appear as vertices of the one-loop fiber for MHV. With the set of vertices at hand, it is simple to write the corresponding canonical form as a sum over vertices. 
\subsection{One-Loop Geometry} 
The loop momentum amplituhedron $\mathcal{M}_{n,k,L}$ was defined in \cite{Ferro:2022abq}: each point in $\mathcal{M}_{n,k,L}$ is specified by a point $(\lambda,\tilde\lambda)$ in the tree-level momentum amplituhedron $\mathcal{M}_{n,k,0}$, together with a collection of $L$ loop momenta $\ell_i$. The canonical form of the loop momentum amplituhedron $\Omega[\mathcal{M}_{n,k,L}]$ conjecturally encodes the $L$-loop integrand of planar $\mathcal{N}=4$ sYM. When translated to the dual momentum space, as in \cite{Ferro:2023qdp}, the loop momentum amplituhedron is specified by points $(\lambda,\mathbf{x})$, where $\mathbf{x}$ is a null polygon $\mathbf{x}\in\mathcal{P}_{n,k}$, together with a collection of loop dual momenta $y_{i}$, for $i=1,\ldots,L$, satisfying additional positivity conditions. We will denote the set of all $(\lambda,\mathbf{x},y_1,\ldots,y_L)$ corresponding to points in the loop momentum amplituhedron by $\widetilde{\mathcal{M}}_{n,k,L}$ and its canonical differential form by $\Omega_{n,k,L}=\Omega\left[\widetilde{\mathcal{M}}_{n,k,L}\right]$. 

Let us focus on the one-loop momentum amplituhedron $\smash{\mathcal{M}_{n,k,1}}$, which consists of points $\smash{(\lambda,\tilde\lambda,y)}$, where $(\lambda,\tilde\lambda)\in\mathcal{M}_{n,k,0}$. The spinor helicity variables $(\lambda,\tilde\lambda)$ define a null-polygon $\mathbf{x}\equiv\mathbf{x}_{(\lambda,\tilde\lambda)}$ through equation \eqref{eq:lambda-to-x}, and $y$ is required to satisfy that
\begin{itemize}
	\item the distances between $y$ and all vertices $x_a$ of the null polygon are non-negative
	\begin{equation}\label{eq:dist-non-neg}
		(y-x_a)^2\geq 0 \qquad \text{for all } a=1,\ldots,n\,,
	\end{equation}
	\item the sequences
	\begin{equation}\label{eq:signflipsforloops}
		\{\langle a\,a+1\rangle(y-\ell^*_{a\,a+1})^2,\langle a\,a+2\rangle(y-\ell^*_{a\,a+2})^2,\ldots,\langle a\,a+n-1\rangle(y-\ell^*_{a\,a+n-1})^2\}\,,
	\end{equation} 
	have $k$ sign flips for all $a=1,\ldots,n$. 
\end{itemize}
The notion of \emph{one-loop fiber} arises from considering a projection map from $\smash{\widetilde{\mathcal{M}}_{n,k,1}}$ to $\widetilde{\mathcal{M}}_{n,k,0}$ which sends $(\lambda,\mathbf{x},y)\mapsto (\lambda,\mathbf{x})$. The fiber attached to a point $(\lambda,\mathbf{x})\in \widetilde{\mathcal{M}}_{n,k,0}$ is the preimage of this projection map. It consists of all the points $y\in\mathbb{R}^{2,2}$ which satisfy the above positivity and sign-flip definition for fixed $(\lambda,\mathbf{x})$. We denote this one-loop fiber as $\Delta(\mathbf{x})$ (or $\Delta_{n,k}(\mathbf{x})$ if we want to keep track of $n$ and $k$ explicitly). In other words, we associate to a fixed null polygon $\mathbf{x}\in\mathcal{P}_{n,k}$, a subset $\Delta(\mathbf{x})\subset \mathbb{R}^{2,2}$. It is clear that the one-loop fiber encodes the loop-level structure whereas $\widetilde{\mathcal{M}}_{n,k,0}$ encodes the tree-level structure. This thus gives an interpretation of the loop momentum amplituhedron as a \emph{fibration} over the tree-level momentum amplituhedron in the dual momentum space.

It was argued in \cite{Ferro:2023qdp} that the one-loop fiber $\Delta(\mathbf{x})$ is a compact region of $\mathbb{R}^{2,2}$ with facets given by $(y-x_a)^2=0,\, a=1,2,\ldots,n$, \emph{i.e.} it is `cut out' by the null-cones centered at the points $x_a$. For this reason we will often refer to the fiber geometry as a {\it null-cone geometry}. Furthermore, $\Delta(\mathbf{x})$ is a curvy version of a {\it simple polytope} in four dimensions, since all vertices of $\Delta(\mathbf{x})$ have exactly four incident edges. Moreover, the boundary stratification of $\Delta(\mathbf{x})$ can be directly related to familiar notions from the scattering process. The facets $(y-x_a)^2=0$ correspond to forward limits, and lower-dimensional boundaries correspond to cutting the integrand multiple times. In particular, the vertices of $\Delta(\mathbf{x})$ correspond to a subset of maximal cuts of the one-loop integrand which can be split into two groups: 
\begin{itemize}
	\item Composite cuts correspond to the vertices $x_{a}$ defining the null-polygon. They are the solutions to cutting three consecutive propagators
	\begin{equation}
		(y-x_{a-1})^2=(y-x_a)^2=(y-x_{a+1})^2=0\,,
	\end{equation}
	together with the vanishing of the Jacobian for these equations
	\begin{equation}
		\text{Jac}_{(y-x_{a-1})^2=(y-x_a)^2=(y-x_{a+1})^2}=0\,.
	\end{equation}
	\item Quadruple-cut points $q_{abcd}^{\pm}\in \mathbb{R}^{2,2}$ are the two solutions to the quadruple-cut conditions
	\begin{equation}
		(y-x_a)^2=(y-x_b)^2=(y-x_c)^2=(y-x_d)^2=0\,.
	\end{equation}
\end{itemize}

For a fixed null-polygon $\mathbf{x}$ it is straightforward to determine which vertices are inside the one-loop fiber $\Delta(\mathbf{x})$, as it is sufficient to check the positivity conditions \eqref{eq:dist-non-neg} and the sign-flip conditions \eqref{eq:signflipsforloops} for each of the quadruple-cut solutions $q_{abcd}^\pm$. Some care must be taken when counting sign flips, however, since the quadruple-cut points are defined to be null separated to four of the $x_a$. Therefore, some entries in the sequence \eqref{eq:signflipsforloops} will be zero. To remedy this, we simply consider a slight deviation from the point $q^\pm_{abcd}$. Effectively, this replaces the zeroes in the sign-flip sequence by either $\pm 1$. If there exists a replacement which satisfies the correct number of sign flips, then the point $q^\pm_{abcd}$ is a vertex of the one-loop fiber geometry. 

For MHV amplitudes, which are the main interest of this paper, one finds that for all null-polygons $\mathbf{x}\in\mathcal{P}_{n,2}$ the quadruple-cut points that are vertices of $\Delta(\mathbf{x})$ are $\ell^*_{ab}=q_{aa+1bb+1}^+$ for $|a-b|>1$, and all other quadruple-cut points sit outside.

\subsection{Canonical Forms for One-Loop Fibers}\label{sec:chamber_forms}
As explained in \cite{Ferro:2023qdp}, to write down the canonical differential form of the one-loop fiber $\Delta(\mathbf{x})$, it is sufficient to find all quadruple-cut points $q^\pm_{abcd}$ that are inside. This originates from the fact that the one-loop fibers are curvy versions of simple polytopes. Each vertex $q_{abcd}^\pm$ inside $\Delta(\mathbf{x})$ contributes to its canonical form the following expression
\begin{align}\label{eq:combination}
\omega_{abcd}^\pm=\frac{1}{2}(\omega_{abcd}^\square\pm \omega_{abcd}),
\end{align}
where the box integrand is defined as
\begin{align}\label{eq:box-integrand}
\omega_{abcd}^\square&=\pm \dd\log\frac{(y-x_a)^2}{(y-q^\pm_{abcd})^2}\wedge  \dd\log\frac{(y-x_b)^2}{(y-q^\pm_{abcd})^2}\wedge  \dd\log\frac{(y-x_c)^2}{(y-q^\pm_{abcd})^2} \wedge  \dd\log\frac{(y-x_d)^2}{(y-q^\pm_{abcd})^2},\\
&=\frac{4\Delta (x_a-x_c)^2(x_b-x_d)^2 \, \dd^4y}{(y-x_a)^2(y-x_b)^2(y-x_c)^2(y-x_d)^2}\,,\qquad \Delta=\sqrt{(1-u-v)^2-4uv},
\end{align}
where we use the standard cross-ratios $\smash{u=\frac{(x_a-x_b)^2(x_c-x_d)^2}{(x_a-x_c)^2(x_b-x_d)^2}}$, $\smash{v=\frac{(x_a-x_d)^2(x_c-x_c)^2}{(x_a-x_c)^2(x_b-x_d)^2}}$, and 
\begin{align}\label{eq:respm}
\omega_{abcd}&= \dd\log(y-x_a)^2\wedge  \dd\log(y-x_b)^2\wedge  \dd\log(y-x_c)^2 \wedge  \dd\log(y-x_d)^2.
\end{align}
By construction, the form $\omega_{abcd}^\pm$ has the desired property of having non-zero residue on {\it only one} of the quadruple-cut points $q_{abcd}^\pm$ with the corresponding residues given by
\begin{equation}\label{eq:resqijkl}
\mathop{\text{Res}}_{y=q^\pm_{abcd}}\omega_{abcd}^\pm=\pm 1\,,\qquad\mathop{\text{Res}}_{y=q^\pm_{abcd}}\omega_{abcd}^\mp=0\,.
\end{equation}
Importantly, the vertices $q_{abcd}^\pm$ are not the only locations where the forms $\omega_{abcd}^\pm$ have non-vanishing residues. Any form with at least two consecutive indices $\omega^\pm_{aa+1b_1b_2}$ has a non-zero composite residue when taking $(y-x_a)^2=(y-x_{a+1})^2=(y-x_{b_p})^2=0$ together with cutting the corresponding Jacobian. In particular, the forms with three consecutive indices $\omega^\pm_{a-1aa+1b}$ have a composite residue at the point $x_a$ which evaluates to
\begin{equation}\label{eq:resx}
\mathop{\text{Res}}_{x=x_a}\omega_{a-1aa+1b}^\pm=\frac{1}{2}\,.
\end{equation}
The one-loop fiber $\Delta(\mathbf{x})$ always contains exactly one vertex of the form $q^+_{a-1aa+1b}$ and one vertex of the form $q^-_{a-1aa+1c}$ for all $a=1,\ldots,n$, from which follows that the canonical form of $\Delta(\mathbf{x})$ has unit residue at each of the vertices $x_a$.

With these definitions for $\omega_{abcd}^\pm$,  the canonical form for the one-loop fiber can be written as a sum over the set of all quadruple-cut points $\mathcal{V}(\Delta(\mathbf{x}))$ inside $\Delta(\mathbf{x})$
\begin{align}\label{eq:form-as-sum}
	\Omega\left[\Delta(\mathbf{x})\right]=\sum_{q^\pm_{abcd}\in\mathcal{V}(\Delta(\mathbf{x}))} \text{sgn}^\pm_{abcd}\; \omega^\pm_{abcd}\,,
\end{align}
where the signs $\text{sgn}^\pm_{abcd}$ are fixed by demanding projective invariance of $\Omega\left[\Delta(\mathbf{x})\right]$ under the following rescaling:
\begin{equation}
(y-p)^2\to \Lambda(y)(y-p)^2\,,
\end{equation}
for all vertices $p$ of $\Delta({\bf x})$. If we fix the ordering of $(a,b,c,d)$ to follow the standard ordering on $1,2,\ldots,n$, then we find $\text{sgn}_{abcd}=1$ for all $q^\pm_{abcd}$.

For MHV amplitudes, the canonical form of $\Delta(\mathbf{x})$ can be written very explicitly since the only quadruple-cut points inside $\Delta(\mathbf{x})$ are $\ell^*_{ab}=q_{aa+1bb+1}^+$ for $|a-b|>1$.  
%
%
This allows us to write the canonical form of the full one-loop MHV momentum amplituhedron as
\begin{align}\label{eq:one-loop-integrand-chamber-sum}
	\Omega_{n,2,1}=\Omega_{n,2,0}\wedge \Omega[\Delta_{n,2}]\,,
\end{align}
where the canonical form of the one-loop fiber is given by
\begin{align}
	\Omega[\Delta_{n,2}] = \sum_{|a-b|>1} \omega^+_{aa+1bb+1} \,,
\end{align}
and the distance $|a-b|$ is taken with cyclic boundary conditions.

\section{Two Loops}
\label{sec:two_loops}
In this section we move on to study the two-loop MHV integrands for which the extension of the formalism from the previous sections will lead to new formulae based on geometry. We begin by introducing a conceptual framework we refer to as {\it fibrations of fibrations}. The main idea in this approach is to view the two-loop geometry as an iterated fibration over both the tree-level kinematics given by a null polygon ${\bf x}$ and the position of the loop momentum $y_1$ within the one-loop fiber $\Delta(\mathbf{x})$. As we shall see this will lead to a natural decomposition of the one-loop fiber into one-loop chambers. The significance of the one-loop chambers is that their corresponding {\it two-loop fibers}, to be defined in the next section, are combinatorially equivalent. Ultimately, this will lead to a factorised form for the two-loop MHV integrand analogous to that of \eqref{eq:one-loop-integrand-chamber-sum}. 
 
Having introduced the idea of fibrations of fibrations we demonstrate how it works in practice for the five-point example which is simple enough to be presented in full detail. Following this we present a full classification of one-loop chambers for general $n$. By making use of the results of \cite{Parisi:2021oql}, we show how each chamber is characterised by the signs of the distances $(y_1-\ell^*_{ab})^2$, and how this information can be succinctly encoded by a permutation. We then move on to discuss how to compute the canonical forms for both the one-loop chambers and their corresponding two-loop fibers as a sum over vertices. The main result will be the two-loop MHV integrand written explicitly in the fibration of fibration form \eqref{eq:fibrations}. 
 
\subsection{Fibration of Fibration}
At one-loop the key insight of \cite{Ferro:2023qdp} was that splitting the space of null polygons $\mathcal{P}_{n,k}$ into tree-level chambers leads to an expression for the one-loop integrand \eqref{eq:intro-factor} where each term takes a factorised form. We now wish to show how this can be extended to two-loops for all MHV integrands.

At two loops the momentum amplituhedron $\widetilde{\mathcal{M}}_{n,k,2}$ is parametrised by a null polygon $\mathbf{x}\in\mathcal{P}_{n,2}$, and by two loop momenta  $y_1$ and $y_2$, both constrained to the one-loop fiber $\Delta({\bf x})$, with the additional mutual positivity condition $(y_1-y_2)^2 \geq 0$. We begin by fixing the null polygon $\mathbf{x}$ together with $y_1 \in \Delta({\bf x})$ and define the two-loop fiber $\Delta({\bf x},y_1)$ to be:
\begin{align}
\Delta({{\bf x}},y_1):=\{ y \in \Delta({{\bf x}}) \ | \ (y-y_1)^2 \geq 0 \}.
\end{align}
The combinatorial structure of $\Delta({{\bf x}},y_1)$ will differ as we vary $y_1 \in \Delta({{\bf x}})$ and, in analogy to the situation at one loop, we wish to consider the equivalence classes of points for which the combinatorial structure of $\Delta({{\bf x}},y_1)$ remains unchanged. We refer to these equivalence classes as one-loop chambers which we label as $\mathfrak{c}^{\{ i \} }_{n,2}$, where the superscript indicates the one-loop chamber in which the first loop momentum $y_1$ resides. This will allow us to write the two-loop MHV integrand in the form of a {\it fibration of fibration} as 
\begin{align}
\Omega_{n,2,2} =\Omega_{n,2,0} \wedge \sum_{i} \Omega\left[\mathfrak{c}_{n,2}^{\{ i \}}\right] \wedge \Omega\left[\Delta^{\{i\}}_{n,2}\right],
\label{eq:fibrations}
\end{align}
where $\Delta^{\{i\}}_{n,2}=\Delta(\mathbf{x},y_1)$ is the two-loop fiber geometry for $y_1\in\mathfrak{c}_{n,2}^{\{ i \}}$.
Here the sum is over all one-loop chambers and the last factor is the canonical form for the corresponding two-loop fiber. Remarkably, as we will see in the coming sections, the two-loop fibers for MHV amplitudes are again curvy versions of simple polytopes, and as such their canonical forms can be calculated via a sum over their vertices
\begin{align}
 \Omega\left[\Delta^{\{i\}}_{n,2}\right] = \sum_{v \in \mathcal{V}_{n,2}^{\{i\}}} \omega_v,
\end{align}
where here we have defined $\mathcal{V}_{n,2}^{\{ i\}}$ to be the vertex set of $\Delta^{\{i\}}_{n,2}$.
\subsection{Warm-up Example}
\label{sec:warmup}
\begin{figure}
\center
\begin{tikzpicture}[scale=2]
\foreach \a in {1,2,...,5} { 
\coordinate (out\a) at ($(\a*72+18:2.5cm)$);
\coordinate (in\a) at ($(\a*72+18+180:1.2cm)$);
\coordinate (l\a) at ($(\a*72+12:1.7cm)$);
\coordinate (r\a) at ($(\a*72+24:1.7cm)$);
}
\draw[thick] (out1)--(out2)--(out3)--(out4)--(out5)--(out1);
\draw[thick,red] (out1) -- (in4)  node[sloped,pos=0.5,allow upside down,red]{\arrowIn};
\draw[thick,red]  (in3) -- (out1) node[sloped,pos=0.5,allow upside down,red]{\arrowIn};
\draw[thick,red] (out5) -- (in3)  node[sloped,pos=0.5,allow upside down,red]{\arrowIn};
\draw[thick,red] (in2) -- (out5)  node[sloped,pos=0.5,allow upside down,red]{\arrowIn};
\draw[thick,red] (out4) -- (in2)  node[sloped,pos=0.5,allow upside down,red]{\arrowIn};
\draw[thick,red] (in1) -- (out4)  node[sloped,pos=0.5,allow upside down,red]{\arrowIn};
\draw[thick,red] (out3) -- (in1)  node[sloped,pos=0.5,allow upside down,red]{\arrowIn};
\draw[thick,red] (in5) -- (out3)  node[sloped,pos=0.5,allow upside down,red]{\arrowIn};
\draw[thick,red] (out2) -- (in5)  node[sloped,pos=0.5,allow upside down,red]{\arrowIn};
\draw[thick,red] (in4) -- (out2)  node[sloped,pos=0.5,allow upside down,red]{\arrowIn};
\draw[thick,red] (in3) -- (in1)  node[sloped,pos=0.5,allow upside down,red]{\arrowIn};
\draw[thick,red] (in5) -- (in3)  node[sloped,pos=0.5,allow upside down,red]{\arrowIn};
\draw[thick,red] (in2) -- (in5)  node[sloped,pos=0.5,allow upside down,red]{\arrowIn};
\draw[thick,red] (in4) -- (in2)  node[sloped,pos=0.5,allow upside down,red]{\arrowIn};
\draw[thick,red] (in1) -- (in4)  node[sloped,pos=0.5,allow upside down,red]{\arrowIn};
\foreach \a in {1,2,...,5}{
\draw[fill,blue] (in\a) circle (1pt);
\draw[fill,blue] (out\a) circle (1pt);
}
\foreach \a/\b in {1/1,2/5,3/4,4/3,5/2}{
\filldraw[black] (\a*72+18:2.7cm) circle (0pt) node[]{$x_{\b}$};
}
\filldraw[black] ($1.25*(in2)$) circle (0pt) node[]{$\textcolor{blue}{q^+_{1234}}$};
\filldraw[black] ($1.25*(in1)$) circle (0pt) node[]{$\textcolor{blue}{q^+_{2345}}$};
\filldraw[black] ($1.25*(in5)$) circle (0pt) node[]{$\textcolor{blue}{q^-_{1345}}$};
\filldraw[black] ($1.25*(in4)$) circle (0pt) node[]{$\textcolor{blue}{q^+_{1245}}$};
\filldraw[black] ($1.25*(in3)$) circle (0pt) node[]{$\textcolor{blue}{q^-_{1235}}$};
\filldraw[black] ($1.4*0.5*(in5)+1.4*0.5*(in2)$) circle (0pt) node[]{\textcolor{red}{{\tiny $134$}}};
\filldraw[black] ($1.5*0.5*(in1)+1.5*0.5*(in3)$) circle (0pt) node[]{\textcolor{red}{{\tiny $245$}}};
\filldraw[black] ($1.4*0.5*(in2)+1.4*0.5*(in4)$) circle (0pt) node[]{\textcolor{red}{{\tiny $135$}}};
\filldraw[black] ($1.4*0.5*(in3)+1.4*0.5*(in5)$) circle (0pt) node[]{\textcolor{red}{{\tiny $124$}}};
\filldraw[black] ($1.4*0.5*(in4)+1.4*0.5*(in1)$) circle (0pt) node[]{\textcolor{red}{{\tiny $235$}}};
\filldraw[black] (l1) circle (0pt) node[]{\textcolor{red}{{\tiny $125^+$}}};
\filldraw[black] (r1) circle (0pt) node[]{\textcolor{red}{{\tiny $125^-$}}};
\filldraw[black] (l5) circle (0pt) node[]{\textcolor{red}{{\tiny $123^+$}}};
\filldraw[black] (r5) circle (0pt) node[]{\textcolor{red}{{\tiny $123^-$}}};
\filldraw[black] (l4) circle (0pt) node[]{\textcolor{red}{{\tiny $234^+$}}};
\filldraw[black] (r4) circle (0pt) node[]{\textcolor{red}{{\tiny $234^-$}}};
\filldraw[black] (l3) circle (0pt) node[]{\textcolor{red}{{\tiny $345^+$}}};
\filldraw[black] (r3) circle (0pt) node[]{\textcolor{red}{{\tiny $345^-$}}};
\filldraw[black] (l2) circle (0pt) node[]{\textcolor{red}{{\tiny $145^+$}}};
\filldraw[black] (r2) circle (0pt) node[]{\textcolor{red}{{\tiny $145^-$}}};
\end{tikzpicture}
\caption{One-skeleton of the one-loop fiber geometry for the five-point MHV amplitude. The edges correspond to triple-cut solutions and are therefore labelled by triples $abc$, with $a,b,c=1,\ldots,n$.}
\label{fig:oneskeleton52}
\end{figure}
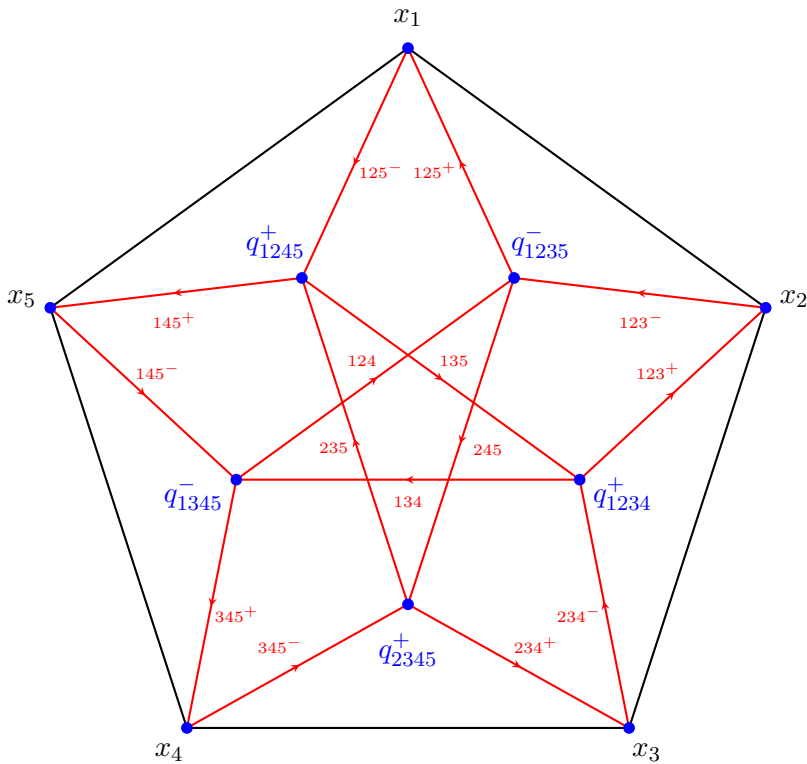

To make the discussion of the previous section more concrete we begin by providing exhaustive details for the two-loop five-point MHV amplitude. As is the case for all MHV amplitudes, there is only one tree-level chamber, meaning that for all possible null polygons ${\bf x} \in \mathcal{P}_{5,2}$ the one-loop fiber geometry $\Delta(\mathbf{x})$ has the same combinatorial structure. As explained in \cite{Ferro:2023qdp}, the one-loop fiber geometry for $n=5$ has 10 vertices: five vertices of the null polygon $x_a$, $a=1,\ldots,5$, together with five quadruple-cut points: 
\begin{equation}\label{eq:qsforn5}
\{q_{1234}^+,q_{2345}^+,q_{3451}^+,q_{4512}^+,q_{5123}^+\}.
\end{equation}
The one-skeleton of this one-loop fiber geometry containing all vertices and edges of the geometry is depicted in Fig.~\ref{fig:oneskeleton52}. 

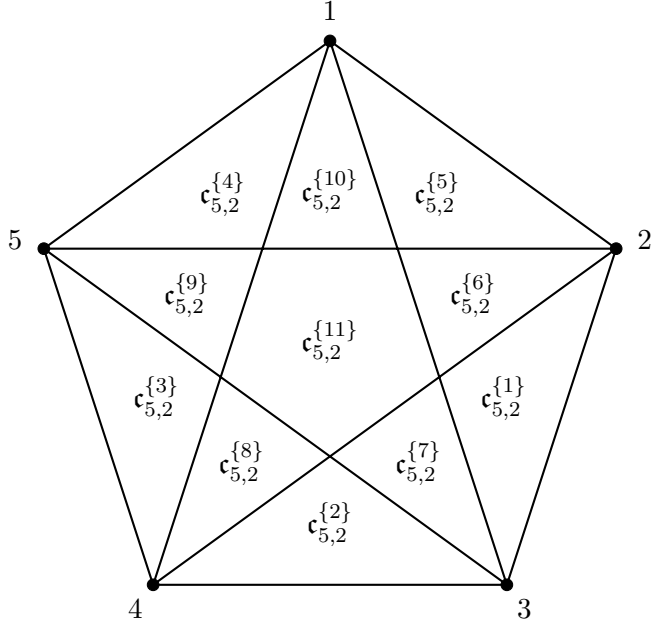
\begin{figure}
\center
\begin{tikzpicture}[scale=2]
\foreach \a in {1,2,...,5} { 
\coordinate (out\a) at ($(\a*72+18:2.cm)$);
}
\draw[thick] (out1)--(out2)--(out3)--(out4)--(out5)--(out1);
\draw[thick] (out1)--(out3)--(out5)--(out2)--(out4)--(out1);
\foreach \a/\b in {1/1,2/5,3/4,4/3,5/2} { 
\filldraw[black] ($1.1*(out\a)$) circle (0pt) node[]{$\b$};
}
\foreach \a in {1,2,...,5}{
\draw[fill,black] (out\a) circle (1.1pt);
}
\filldraw[black] ($(0,0)$) circle (0pt) node[]{$\mathfrak{c}^{\{11\}}_{5,2} $};
\filldraw[black] ($0.5*(out1)$) circle (0pt) node[]{$\mathfrak{c}^{\{10\}}_{5,2} $};
\filldraw[black] ($0.5*(out2)$) circle (0pt) node[]{$\mathfrak{c}^{\{9\}}_{5,2} $};
\filldraw[black] ($0.5*(out3)$) circle (0pt) node[]{$\mathfrak{c}^{\{8\}}_{5,2} $};
\filldraw[black] ($0.5*(out4)$) circle (0pt) node[]{$\mathfrak{c}^{\{7\}}_{5,2} $};x
\filldraw[black] ($0.5*(out5)$) circle (0pt) node[]{$\mathfrak{c}^{\{6\}}_{5,2} $};
\filldraw[black] ($0.75*0.5*(out2)+0.75*0.5*(out3)$) circle (0pt) node[]{$\mathfrak{c}^{\{3\}}_{5,2} $};
\filldraw[black] ($0.75*0.5*(out3)+0.75*0.5*(out4)$) circle (0pt) node[]{$\mathfrak{c}^{\{2\}}_{5,2} $};
\filldraw[black] ($0.75*0.5*(out4)+0.75*0.5*(out5)$) circle (0pt) node[]{$\mathfrak{c}^{\{1\}}_{5,2} $};
\filldraw[black] ($0.75*0.5*(out5)+0.75*0.5*(out1)$) circle (0pt) node[]{$\mathfrak{c}^{\{5\}}_{5,2} $};
\filldraw[black] ($0.75*0.5*(out1)+0.75*0.5*(out2)$) circle (0pt) node[]{$\mathfrak{c}^{\{4\}}_{5,2} $};
\end{tikzpicture}
\caption{One-loop chambers for the five-point MHV amplitude. Each chord $(a,b)$ divides the pentagon into a triangle and a quadrilateral, which correspond to the regions of $\Delta_{5,2}$ for which $(y-\ell^*_{ab})^2>0$ and $(y-\ell^*_{ab})^2<0$, respectively.}
\label{fig:chambers52}
\end{figure}

To study the two-loop geometry, we take a fixed null polygon $\mathbf{x}\in\mathcal{P}_{5,2}$, and we fix $y_1\in\Delta(\mathbf{x})$. The two-loop fiber geometry $\Delta(\mathbf{x},y_1)$ is defined as all points in $\Delta(\mathbf{x})$ that are additionally positively separated from $y_1$. By explicit computation, for example by sampling random points $y_1\in\Delta(\mathbf{x})$, one finds 11 combinatorially inequivalent two-loop fiber geometries, each of which define a one-loop chamber $\mathfrak{c}^{\{ i \}}_{5,2}$ for $i=1,\ldots,11$. These 11 chambers are disjoint and they subdivide the one-loop fiber geometry $\Delta(\mathbf{x})$ into smaller regions and each region is specified by a set of inequalities. It is straightforward to identify the inequalities which define a given chamber as they are governed by the signs of the distances from $y_1$ to the one-loop geometry quadruple-cut points \eqref{eq:qsforn5}. Not all possible combinations of signs are realised for points inside $\Delta(\mathbf{x})$, and it is easy to check that the allowed combinations are those given in Table \ref{tab:n5dist}.
\begin{table}
\center
\begin{tabular}{c||c|c|c|c|c||c|c|c|c|c||c}
&\small $\mathfrak{c}_{5,2}^{\{1\}}$&\small$\mathfrak{c}_{5,2}^{\{2\}}$&\small$\mathfrak{c}_{5,2}^{\{3\}}$&\small$\mathfrak{c}_{5,2}^{\{4\}}$&\small$\mathfrak{c}_{5,2}^{\{5\}}$&\small$\mathfrak{c}_{5,2}^{\{6\}}$&\small$\mathfrak{c}_{5,2}^{\{7\}}$&\small$\mathfrak{c}_{5,2}^{\{8\}}$&\small$\mathfrak{c}_{5,2}^{\{9\}}$&\small$\mathfrak{c}_{5,2}^{\{10\}}$&\small$\mathfrak{c}_{5,2}^{\{11\}}$\\\hline\hline
\small$\ell^*_{13}=q_{1234}^+$&+&$-$&$-$&$-$&$+$&+&$-$&$-$&$-$&$-$&$-$\\\hline
\small$\ell^*_{24}=q_{2345}^+$&$+$&+&$-$&$-$&$-$&$-$&$+$&$-$&$-$&$-$&$-$\\\hline
\small$\ell^*_{35}=q_{3451}^+$&$-$&$+$&+&$-$&$-$&$-$&$-$&$+$&$-$&$-$&$-$\\\hline
\small$\ell^*_{14}=q_{4512}^+$&$-$&$-$&$+$&+&$-$&$-$&$-$&$-$&$+$&$-$&$-$\\\hline
\small$\ell^*_{25}=q_{5123}^+$&$-$&$-$&$-$&$+$&$+$&$-$&$-$&$-$&$-$&$+$&$-$\\
\end{tabular}
\caption{The collections of signs of distances $(y_1-v)^2$ to the quadruple-cut points $v\in\{q_{1234}^+,q_{2345}^+,q_{3451}^+,q_{4512}^+,q_{5123}^+\}$ for all one-loop chambers for $n=5$, $k=2$.}
\label{tab:n5dist}
\end{table}
In this case the 11 chambers and their adjacency can be illustrated\footnote{The fact that all chambers can be depicted and labelled as in Fig.~\ref{fig:chambers52} is only possible for $n=5$, since in this case, the one-loop chamber decomposition is dual to the chamber decomposition of a convex pentagon. Such duality however does not generalise to higher $n$.
} as in Fig.~\ref{fig:chambers52}. 

As illustrated by Table~\ref{tab:n5dist}, the one-loop chambers can be divided into three cyclic classes, and the one-loop geometry is the union of all chambers 
\begin{align}
\Delta(\mathbf{x})=  \left( \mathfrak{c}^{\{1\}}_{{5,2}}\cup  \ldots \cup\mathfrak{c}^{\{5 \}}_{5,2} \right) \cup \left( \mathfrak{c}^{\{6\} }_{5,2}\cup  \ldots \cup\mathfrak{c}^{\{10 \}}_{5,2} \right) \cup \left( \mathfrak{c}^{\{11\} }_{5,2} \right)\,.
\end{align}
For each one-loop chamber $\mathfrak{c}^{\{i \}}_{5,2} $ for $i=1,\ldots,11$, one can study the second loop fiber geometry $\Delta_{5,2}^{\{i\}}$,  and in particular find all of its vertices. The vertices of any two-loop fiber geometry form a subset of the following points:
\begin{itemize}
\item vertices of the null polygon $x_a$,
\item quadruple-cut points involving four vertices of the null polygon: $q_{aa+1bb+1}^+$,
\item quadruple-cut points involving three vertices of the null polygon and $y_1$: $q_{aa+1by_1}^\pm$.
\end{itemize}
From explicit calculations in each chamber, we find that the vertex sets are:
\begin{align}
\mathcal{V}_{5,2}^{\{1\}}&=\{x_a,q_{1234}^+,q_{2345}^+,q_{123y_1}^-,q_{125y_1}^+,q_{125y_1}^-,q_{145y_1}^+,q_{145y_1}^-,q_{345y_1}^+,q_{245y_1}^-,q_{235y_1}^+,q_{124y_1}^+,q_{134y_1}^-\}\,,\nonumber\\
\mathcal{V}_{5,2}^{\{6\}}&=\{x_a,q_{1234}^+,q_{123y_1}^-,q_{125y_1}^+,q_{125y_1}^-,q_{145y_1}^+,q_{145y_1}^-,q_{345y_1}^+,q_{345y_1}^-,q_{234y_1}^+,q_{124y_1}^+,q_{134y_1}^-\}\,,\nonumber\\
\mathcal{V}_{5,2}^{\{11\}}&=\{x_a,q_{123y_1}^+,q_{123y_1}^-,q_{125y_1}^+,q_{125y_1}^-,q_{145y_1}^+,q_{145y_1}^-,q_{345y_1}^+,q_{345y_1}^-,q_{234y_1}^+,q_{234y_1}^-\}\,,\label{eq:two-loopvertices}
\end{align}
with the remaining chambers obtained by cyclic rotation of labels.
Moreover, every vertex is adjacent to exactly four edges, making the two-loop fiber geometry a curvy version of a simple polytope.

Following the same logic we applied to the one-loop problem, we can find the explicit formula for the canonical form of the two-loop fiber geometry, which after summing over all chambers gives the following expression for the two-loop integrand:
\begin{align}\label{eq:omega252}
\Omega_{5,2,2} = \Omega_{5,2,0}\wedge\sum_{i=1}^{11} \left(\Omega[\mathfrak{c}_{5,2}^{\{ i \}}] \wedge \Omega[\Delta^{\{ i\}}_{5,2}]\right).
\end{align}
Using \eqref{eq:two-loopvertices}, the three cyclic types of chambers contribute
\begin{align}
\Omega[\Delta_{5,2}^{\{1\}}]  &=\omega^{\Box}_{145y_1}+ \omega^{\Box}_{125y_1}+\omega^-_{123y_1}+\omega^+_{345y_1}+\omega^+_{1234}+\omega^+_{124y_1}+ \omega^-_{134y_1} +\omega^+_{2345}+\omega^-_{245y_1}  +\omega^+_{235y_1} \notag,  \notag \\
\Omega[\Delta_{5,2}^{\{6\}}]  &=  \omega^{\Box}_{145y_1}+ \omega^{\Box}_{125y_1}+\omega^-_{123y_1}+\omega^+_{234y_1}+\omega^{\Box}_{345y_1}+\omega^+_{1234}+\omega^+_{124y_1}+\omega^-_{134y_1},  \notag \\
  \Omega[\Delta_{5,2}^{\{11\}}]  &=  \omega^{\Box}_{145y_1}+\omega^{\Box}_{125y_1}+\omega^{\Box}_{123y_1}+ \omega^{\Box}_{234y_1}+\omega^{\Box}_{345y_1},
\label{eq:cham_5}
\end{align}
where we recall that $\omega_{abcd}^{\Box}=\omega_{abcd}^++\omega_{abcd}^-$, and the formulae for the remaining eight chambers can be obtained by cyclic rotation of labels. Notice that all signs in front of the differential forms are positive. Understanding the structure of the above answer will lead to a straightforward generalisation to all MHV$_n$ two-loop amplitudes. 
Interestingly, there exists an alternative expression for the canonical forms of the two-loop fiber geometries as a sum over box and chiral pentagon differential forms $\cpp_{aa+1bb+1x}$ defined as
\begin{align}
\cpp_{aa+1bb+1x} = \frac{4 s_{ab} (y-q^-_{aa+1bb+1})^2(x-q^+_{aa+1bb+1})^2}{(y-x_a)^2(y-x_{a+1})^2(y-x_b)^2(y-x_{b+1})^2(y-x)^2}\dd^4 y\,,
\end{align}
where 
$x^\mu$ is an arbitrary point in dual-momentum space. We can relate the chiral pentagon to the non-chiral pentagon and box differential forms via
\begin{align}\label{eq:chiralpenttoboxes}
	{\cpp_{aa+1bb+1x} } &=  \frac{1}{2}\left(\ncp_{aa+1bb+1x}\pm\omega^{\Box}_{aa+1bb+1}  \pm\omega^{\Box}_{aa+1bx}  \pm\omega^{\Box}_{aa+1b+1x}  \pm\omega^{\Box}_{bb+1ax}  \pm\omega^{\Box}_{bb+1a+1x} \right)\,,
\end{align}
where the non-chiral pentagon is given by
\begin{align}
	\ncp_{abcde} = \omega_{abcd}-\omega_{abce}+\omega_{abde}-\omega_{acde}+\omega_{bcde}\,,
\end{align}
and the signs in front of box differential forms depend on $x$.
The latter originates from the fact that when any two indices of the box differential form are consecutive, namely any two points are null separated, then the box differential form simplifies to
\begin{align}
	\omega^\Box_{aa+1bc} = \frac{4\, |(x_a-x_b)^2(x_{a+1}-x_c)^2- (x_a-x_c)^2(x_{a+1}-x_b)^2|}{(y-x_a)^2(y-x_{a+1})^2(y-x_b)^2(y-x_c)^2}\dd^4y\,.
\end{align}
Then, the signs in formula \eqref{eq:chiralpenttoboxes} are related to the signs of the expressions inside the absolute value.

In the basis of chiral pentagons and boxes, the canonical forms for the two-loop fibers become
\begin{align}
\Omega[\Delta_{5,2}^{\{1\}}] &=\omega^{\Box}_{123y_1}+\omega^{\Box}_{234y_1}+\omega^{\Box}_{345y_1}+\omega^{\Box}_{451y_1}+\omega^{\Box}_{512y_1}+\cpp_{1234y_1}+\cpp_{2345y_1}, \notag \\
\Omega[\Delta_{5,2}^{\{6\}}] &= \omega^{\Box}_{123y_1}+\omega^{\Box}_{234y_1}+\omega^{\Box}_{345y_1}+\omega^{\Box}_{451y_1}+\omega^{\Box}_{512y_1}+\cpp_{1234y_1},\notag \\
\Omega[\Delta_{5,2}^{\{11\}}] &= \omega^{\Box}_{123y_1}+\omega^{\Box}_{234y_1}+\omega^{\Box}_{345y_1}+\omega^{\Box}_{451y_1}+\omega^{\Box}_{512y_1},
\label{eq:cham_forms_cp}
\end{align}
where we used the fact that the signs in formula \eqref{eq:chiralpenttoboxes} are fixed when $y_1$ is inside a one-loop chamber.
Comparing this to the sign vectors for each chamber in Table \ref{tab:n5dist}, one notices that the chiral pentagons $\cpp_{aa+1bb+1y_1}$ which appear for a given one-loop chamber are exactly those for which the corresponding vertex $q^+_{aa+1bb+1}$ is positively separated from the loop momentum $y_1$.

Surprisingly, also the canonical differential forms of the one-loop chambers can be found using a very similar method since the one-loop chambers themselves are curvy versions of simple polytopes and therefore the canonical forms can again be found as a sum over their vertices. We find 
\begin{align}
\Omega[\mathfrak{c}_{5,2}^{\{1\}}]&= \omega^\Box_{\ell^*_{13}x_1x_5\ell^*_{24}},\notag \\
\Omega[\mathfrak{c}_{5,2}^{\{6\}}]&=\omega^-_{x_1\ell^*_{24}\ell^*_{25}x_4}- \omega^+_{\ell^*_{13}x_1\ell^*_{25}x_4}- \omega^-_{\ell^*_{13}x_1\ell^*_{24}x_4}+ \omega^-_{\ell^*_{13}x_1\ell^*_{24}\ell^*_{25}}+ \omega^+_{\ell^*_{13}\ell^*_{24}\ell^*_{25}x_4}=\cpp_{\ell^*_{13}x_1\ell^*_{24}x_4 \ell^*_{25}},\notag \\
\Omega[\mathfrak{c}_{5,2}^{\{11\}}]&=\omega^-_{\ell^*_{13}\ell^*_{14}\ell^*_{25}\ell^*_{35}}+\omega^-_{\ell^*_{24}\ell^*_{25}\ell^*_{13}\ell^*_{14}}+\omega^-_{\ell^*_{35}\ell^*_{13}\ell^*_{24}\ell^*_{25}}+\omega^-_{\ell^*_{14}\ell^*_{24}\ell^*_{35}\ell^*_{13}}+\omega^-_{\ell^*_{25}\ell^*_{35}\ell^*_{14}\ell^*_{24}}.
\label{eq:cham_forms_5}
\end{align}
Interestingly, the canonical form for $\mathfrak{c}_{5,2}^{\{11\}}$ equals the five-point $\overline{\text{MHV}}_5$ amplitude for the null-polygon formed of all $\ell^*_{aa+2}$.

Before proceeding to general $n$, we  provide an interesting way in which we can re-organise formula \eqref{eq:omega252}. If one rewrites \eqref{eq:omega252} with respect to the last entry in each wedge product, one finds the coefficients that multiply contributions coming from any single vertex of the two-loop fiber geometry, namely
\begin{align}
\frac{\Omega_{5,2,2} }{\Omega_{5,2,0}} &=\left(\Omega^{(1)}_{12345} - \Omega^{(1)}_{451 \ell^*_{13}} \right)\omega_{123y_1}^+ +\left( \Omega^{(1)}_{12345} - \Omega^{(1)}_{345 \ell^*_{25}}\right)\omega_{123y_1}^- +\left( \Omega^{(1)}_{451\ell^*_{13}} \right)\omega_{1234}^+ + \text{ cyclic} \notag \\
&\pm\left(  \Omega^{(1)}_{451 \ell^*_{13}}  -\Omega^{(1)}_{234\ell^*_{14}}  \right)\omega_{124y_1}^\pm \pm\left(  \Omega^{(1)}_{345 \ell^*_{25}}  -\Omega^{(1)}_{512\ell^*_{24}}  \right)\omega_{235y_1}^\pm\pm\left(  \Omega^{(1)}_{451 \ell^*_{13}}  -\Omega^{(1)}_{123\ell^*_{35}}  \right)\omega_{134y_1}^\pm \notag \\
&\pm\left(  \Omega^{(1)}_{234 \ell^*_{14}}  -\Omega^{(1)}_{512\ell^*_{24}}  \right)\omega_{245y_1}^\pm\pm\left(  \Omega^{(1)}_{345 \ell^*_{25}}  -\Omega^{(1)}_{123\ell^*_{35}}  \right)\omega_{135y_1}^\pm,
\label{eq:two_loop_osc_5}
\end{align}
where $\Omega_{5,2,2}/\Omega_{5,2,0}$ means the canonical form of the two loop momentum amplituhedron $\widetilde{\mathcal{M}}_{5,2,2}$ with the tree-level part removed, and the signs $\pm$ on the forms depend on the explicit position of $y_1$. 
Here we have defined $\Omega^{(1)}_{a \ldots b \ell^*_{a-1b}}$ to be the one-loop MHV integrand (without the tree-level part) evaluated on the points $\{ x_a, \ldots, x_b,\ell^*_{a-1b} \}$. Again, it is possible to simplify the formula \eqref{eq:two_loop_osc_5} if we use chiral pentagon differential forms to get
\begin{align}\label{eq:leadingsing5}
\frac{\Omega_{5,2,2} }{\Omega_{5,2,0}} &=\Omega^{(1)}_{12345} \wedge\sum_{a=1}^5\omega_{a-1aa+1y_1}^{\Box}+ \Omega^{(1)}_{451\ell^*_{13}} \wedge\cpp_{1234y_1}+\ \Omega^{(1)}_{512\ell^*_{24}} \wedge\cpp_{2345y_1}\notag\\&+ \Omega^{(1)}_{123\ell^*_{35}} \wedge\cpp_{3451y_1}+\Omega^{(1)}_{234\ell^*_{14}} \wedge\cpp_{4512y_1}+ \Omega^{(1)}_{345\ell^*_{25}} \wedge\cpp_{5123y_1}.
\end{align}
As we will see in the next section, there exists a natural generalisation of this formula for all $n$.

\subsection{One-Loop Chambers} 
As we have already seen in the five-point example, the structure of the two-loop fiber geometry is combinatorially identical for regions with fixed signs for the distances of $y_1$ to each vertex $\ell^*_{ab}$ of the one-loop fiber. This fact generalizes to all two-loop fiber geometries for MHV amplitudes. Therefore, in order to determine the set of all one-loop chambers, we only need to determine the set of all possible sign patterns within the one-loop fiber geometry for the following table of distances
\begin{align}
\begin{matrix}
\text{sgn}_{12}&\text{sgn}_{13}&\ldots&\text{sgn}_{1n}\\
&\text{sgn}_{23}&\ldots&\text{sgn}_{2n} \\
& &\ddots&\vdots& \\
& & & \text{sgn}_{n-1n} 
\end{matrix}
\quad \quad \quad \quad \quad \quad \quad \quad
\begin{matrix*}[l]
\text{sgn}_{ab} = \text{sign}(y_1-q_{aa+1bb+1}^+)^2,\\
\\
\\
\text{sgn}_{aa+1} = \text{sign}(y_1-x_{a+1})^2.
\end{matrix*}
\label{eq:signs}
\end{align}
Thankfully, this problem was already solved in \cite{Parisi:2021oql}, albeit in a slightly different context. The focus of that paper was the decomposition of both the hypersimplex $\widetilde{\Delta}_{k+1,n}$ and the $m=2$ amplituhedron, related to each other using \emph{T-duality} \cite{Lukowski:2020dpn}, into objects coined {\it w-simplices} and {\it w-chambers}, respectively. It was shown how for fixed $k$ and $n$ the $w$-simplices and $w$-chambers are both in one-to-one correspondence with a set of permutations whose cardinality is the Eulerian number $E_{k+1,n-1}$. Moreover, one can read off from these permutations the set of vertices of the corresponding $w$-simplex or, more relevant for us, the set of inequalities which define the corresponding $w$-chamber. Due to the correspondence between the $(m,k',L)=(2,2,0)$ and the $(m,k,L)=(4,2,1)$ amplituhedra, these results can be directly translated to the problem studied here of characterising all one-loop chambers in the dual space. In particular, the one-loop chambers for MHV amplitudes are in one-to-one correspondence with the $w$-simplices inside the hypersimplex $\widetilde{\Delta}_{3,n}$. We now review their construction and describe how to go from a permutation to a sign pattern table \eqref{eq:signs}, emphasising that the content found in the remainder of this section is simply a translation of the results of \cite{Parisi:2021oql}, which we refer to for a more detailed discussion. 

We start by defining the hypersimplex $\widetilde{\Delta}_{k+1,n}$ that is the subset of $\mathbb{R}^n$ defined as
\begin{equation}
\widetilde{\Delta}_{k+1,n}=\{(\tilde{x}_1,\tilde{x}_2,\ldots,,\tilde{x}_n)\in [0,1]^{n}: \tilde{x}_1+\tilde{x}_2+\ldots+\tilde{x}_n=k+1\}.
\end{equation}
The hypersimplex $\widetilde{\Delta}_{k+1,n}$ is therefore an $(n-1)$-dimensional slice of the $n$-dimensional unit cube. Importantly, there exists a particular decomposition of the hypersimplex $\widetilde{\Delta}_{k+1,n}$ into the Eulerian number $E_{k+1,n-1}$ of unit simplices labelled by permutations that we will now describe.
Let $w=w_1\ldots w_n$ be a permutation on $[n]=\{1,2,\ldots,n\}$. Following the discussion of \cite{Parisi:2021oql}, we say that $i$ is a {\it cyclic descent} of $w$ if it appears {\it to the left} of $i-1$ in $w$ where $i-1$ is understood cyclically. That is, $i \in [n]$ is a cyclic descent of $w$ if either $i \neq 1$ and $w^{-1}(i)<w^{-1}(i-1)$ or $i=1$ and $w^{-1}(1)<w^{-1}(n)$. The set of all cyclic descents of a permutation $w$ will be denoted as $\text{cDes}(w)$, and the set of all permutations with $k+1$ cyclic descents and with $w_n = n$ will be denoted by $D_{k+1,n}$. The cardinality of this set provides a counting of the Eulerian numbers $E_{k+1,n-1}$ for general $n$ and $k$, however, we will focus only on the case of $k=2$ which is relevant to the one-loop chambers for MHV amplitudes. 

In order to make the connection to sign patterns, we define the permutation $w^{(r)}$ to be the cyclic rotation of $w \in D_{k+1,n}$ with $w_n=r$, and define the following set 
\begin{align}
I_r(w):=\text{cDes}(w^{(r-1)}).
\end{align}
An example of these sets for sample permutations for $n=5$ and $k=2$ is given in Table \ref{tab:setsI}. With these definitions it is straightforward to start from any permutation $w$ and find a sign pattern in table \eqref{eq:signs} by the following rule
\begin{align}\label{eq:signsdef}
\text{sgn}_{ij} = (-1)^{|I_{i}(w) \cap [i,j-1]|-1} \quad  \text{for } 1\leq i<j \leq n\,,
\end{align}
where $[i,j-1]=\{i,i+1,\ldots,j-1\}$. Applying this rule to the permutations $u$, $v$ and $w$ in Table \ref{tab:setsI} we notice that they correspond to the chambers $\mathfrak{c}^{\{1\}}_{5,2}$, $\mathfrak{c}^{\{6\}}_{5,2}$ and $\mathfrak{c}^{\{11\}}_{5,2}$ respectively.  
\begin{table}[h!]
\begin{center}
\begin{tabular}{c|c|c|c|c|c}
$r$&1&2&3&4&5\\
\hline
$I_r(u)$& $\{1,3,4\}$& $\{2,4,5\}$& $\{3,4,5\}$& $\{1,4,5\}$& $\{1,3,5\}$\\
\hline
$I_r(v)$& $\{1,3,4\}$& $\{2,3,5\}$& $\{3,4,5\}$& $\{2,4,5\}$& $\{1,3,5\}$\\
\hline
$I_r(w)$& $\{1,2,4\}$& $\{2,3,5\}$& $\{1,3,4\}$& $\{2,4,5\}$& $\{1,3,5\}$
\end{tabular}
\end{center}
\caption{An example for constructing the sets $I_{r}$ for the permutations $u=43125$, $v=41325$ and $w=24135$.}
\label{tab:setsI}
\end{table}
Using \eqref{eq:signsdef} for $k=2$ and general $n$ we can define a map from $w$-simplices inside the hypersimplex $\widetilde{\Delta}_{3,n}$ to one-loop chambers inside the one-loop fiber geometry $\Delta_{n,2}$.

\begin{figure}[h!]
\begin{center}
\includegraphics[scale=0.4]{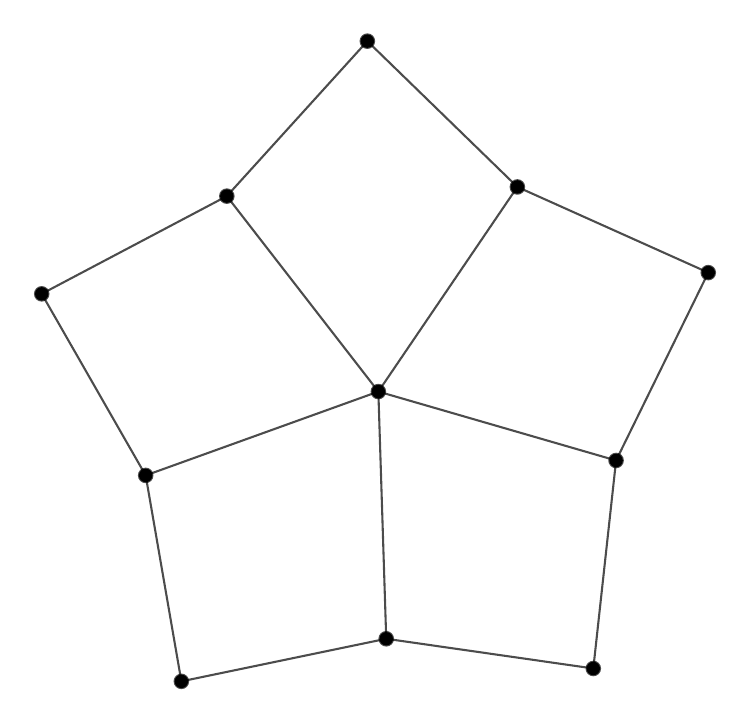}
\includegraphics[scale=0.4]{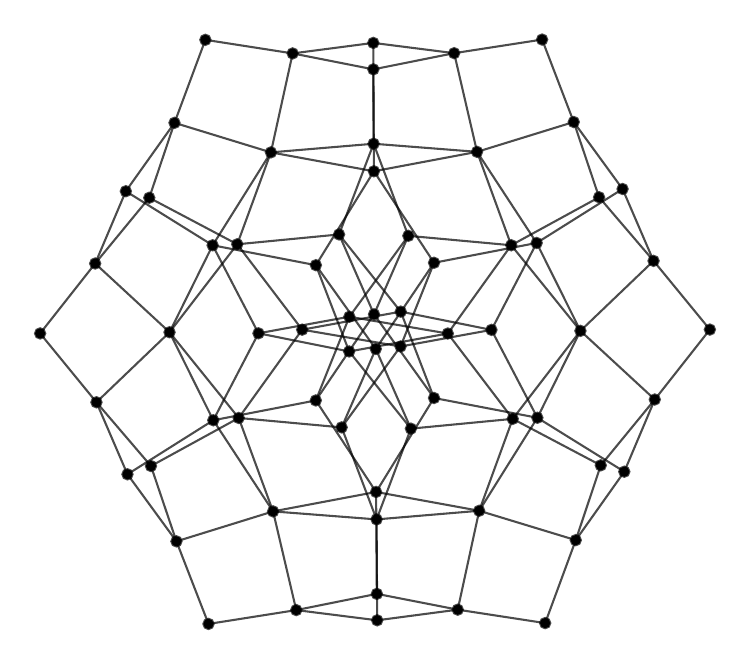}
\includegraphics[scale=0.4]{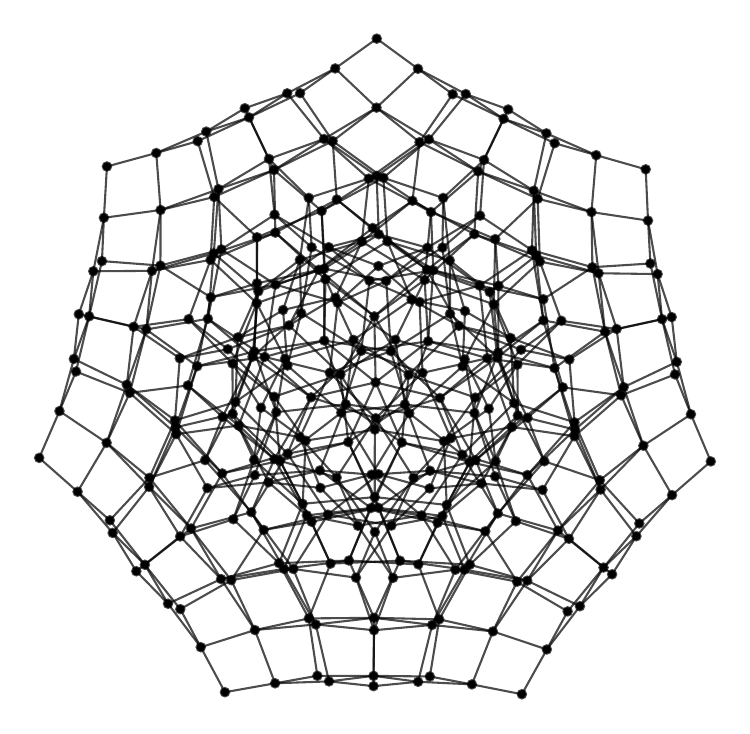}
\end{center}
\caption{The adjacency graph for MHV one-loop chambers for $n=5,6,7$ respectively. Each vertex corresponds to a one-loop chamber and each edge corresponds to a co-dimension one boundary of the form $(y_1-\ell^*_{ab})^2$.}
\label{fig:one_loop_adj}
\end{figure}

As a final remark, with this labelling of the one-loop chambers, we note that it is easy to see whether two chambers are adjacent by simply checking whether (a cyclic representative of) their corresponding permutations differ by a transposition. The adjacency graphs of the one-loop chambers for $n=5,6,7$ are displayed in Figure \ref{fig:one_loop_adj}.

\subsection{Canonical Forms of One-Loop Chambers and Two-Loop Fibers}
In order to find the two-loop canonical forms $\Omega_{n,2,2}$, we need two ingredients for the fibration of fibration idea: the canonical forms of the one-loop chambers $\Omega\left[\mathfrak{c}_{n,2}^{\{i\}}\right]$ and the canonical forms of the two-loop fiber geometries $\Omega\left[\Delta_{n,2}^{\{i\}}\right]$. By exhaustive study of examples for $n<9$, we found that in both cases the geometries have the same property as the one-loop fibers (and the five-point example studied in section \ref{sec:warmup}), namely each vertex is adjacent to exactly four edges. This suggests that in both cases the canonical forms can be found as the sum over contributions coming from vertices. Therefore the task of finding the canonical forms of $\mathfrak{c}_{n,2}^{\{i\}}$ and $\Delta_{n,2}^{\{i\}}$  reduces to the task for finding all their vertices. 

Starting with one-loop chambers, to determine the set of vertices of a given one-loop chamber $\mathfrak{c}_{n,2}^{\{i\}}$, we first need to know its set of co-dimension-one boundaries. There are at least two ways to determine them: one can use the adjacency graphs generated from the adjacency of the $w$-simplices, as in Fig.~\ref{fig:one_loop_adj}; alternatively, there exists a natural map from the set of facets of $w$-simplices to the set of co-dimension-one boundaries of one-loop chambers in the dual space. Using the latter method, one starts by finding all facets of a given $w$-simplex, which can be of the form $\tilde{x}_a=0$, $\tilde{x}_a=1$ or $\tilde{x}_{a+1}+\ldots+ \tilde{x}_{b}=1$. The corresponding one-loop chamber will have the boundaries at positions specified by the following rule:
\begin{center}
\begin{tabular}{c|c}
$\widetilde{\Delta}_{3,n}$&$\widetilde{\mathcal{M}}_{n,2,1}$\\\hline
$\tilde{x}_a=0$& \text{no boundary}\\  
$\tilde{x}_a=1$& $(y_1-x_a)^2=0$\\
$\tilde{x}_{a+1}+\ldots+ \tilde{x}_{b}=1$& $(y_1-\ell^*_{ab})^2=0$
\end{tabular} 
\end{center}
It is clear from this table that the number of co-dimension-one boundaries of any one-loop chamber does not exceed the number of facets of an $(n-1)$-dimensional $w$-simplex. Therefore, for a given $n$, there are maximally $n$ co-dimension-one boundaries for any one-loop chamber.
Knowing the full set of co-dimension-one boundaries, we consider all their possible maximal intersections, namely we study quadruple-cut solutions associated to any four element subset of the boundaries. This prescription produces a set of candidate vertices of a given one-loop chamber. The subset of points which satisfies \eqref{eq:dist-non-neg}, \eqref{eq:signflipsforloops} and the sign pattern conditions for the chamber makes up the set of vertices $\mathcal{V}\left(\mathfrak{c}_{n,2}^{\{i\}}\right)$ of the one-loop chamber $\mathfrak{c}_{n,2}^{\{i\}}$. Then the canonical differential form for a given one-loop chamber can be calculated as a sum over its vertices as
\begin{align}\label{eq:one_loop_fiber_sum_over_vertices}
\Omega[\mathfrak{c}^{\{i\}}_{n,2}]= \sum_{v \in \mathcal{V}(\mathfrak{c}_{n,2}^{\{ i\}})} \omega_v \,.
\end{align}
Using symbolic algebra software, we have found all vertices of one-loop chambers for $n<9$. Currently, we do not have a systematic way to classify them, and therefore we do not present detailed results for one-loop chamber differential forms here. However, as we will see in the next section, after reorganising the results from this section,  a very simple formula can be written for the two-loop integrand, generalising \eqref{eq:leadingsing5}, that relies  on the notion of one-loop leading singularities. 

Having found the canonical forms for one-loop chambers, we can now proceed with two-loop fiber geometries. For a given point $y_1$ in a one-loop chamber $\mathfrak{c}_{n,2}^{\{i\}}$, one notices that the set of  vertices of the corresponding two-loop fiber geometry $\Delta^{\{ i \}}_{n,2}$ is a subset of points $x_a$, $q^+_{aa+1bb+1}$ and $q^\pm_{aa+1by_1}$. Therefore, to determine the vertex set $\mathcal{V}(\Delta_{n,2}^{\{i\}})$ we simply choose a representative point $y_1 \in \mathfrak{c}_{n,2}^{\{i\}}$ and check which points among $\{x_a,q^+_{aa+1bb+1},q^\pm_{aa+1by_1}\}$ simultaneously satisfy the conditions \eqref{eq:dist-non-neg}, \eqref{eq:signflipsforloops} and are non-negatively separated from $y_1$. Knowing the full set of vertices $\mathcal{V}(\Delta_{n,2}^{\{i\}})$ we can write the canonical form as
\begin{align}
\Omega[\Delta^{\{i\}}_{n,2}]= \sum_{v \in \mathcal{V}(\Delta_{n,2}^{\{i\}})} \omega_v,
\label{eq:two_loop_fiber_form_vert}
\end{align}
where for $v=q^+_{aa+1bb+1}$ we take $\omega_v = \omega^+_{aa+1bb+1}$ and for $v=q^\pm_{aa+1by_1}$ we take $\omega_v=\omega^\pm_{aa+1by_1}$. As before, the vertices $x_a$ do not contribute to $\Omega\left[\Delta_{n,2}^{\{i\}}\right]$.

As for the one-loop chambers, one can find the explicit set of vertices contributing to formula \eqref{eq:two_loop_fiber_form_vert} for a fixed two-loop fiber $\Delta_{n,2}^{\{i\}}$. In our case-by-case explorations we found that there exists a very natural way of organising them in terms of box and chiral pentagon contribution as 
\begin{align}
\Omega[\Delta_{n,2}^{\{i\}}] =\sum_{a=1}^n \omega^\Box_{a-1aa+1y_1}+ \sum_{(y_1-\ell^*_{ab} )^2>0} \cpp_{aa+1bb+1y_1},
\label{eq:two_loop_fiber_form_cp}
\end{align}
where the second sum is over all $(a,b)$ for which $\ell^*_{ab}$ is positively separated from $y_1\in\mathfrak{c}_{n,2}^{\{i\}}$. From this expression it straightforward to arrive at the explicit vertex expression in \eqref{eq:two_loop_fiber_form_vert} by using formula \eqref{eq:chiralpenttoboxes}.

With the results for the canonical forms of the one-loop chambers \eqref{eq:one_loop_fiber_sum_over_vertices} and their corresponding two-loop fibers \eqref{eq:two_loop_fiber_form_cp}, we now have all the ingredients that contribute to the two-loop fibration of fibration formula for the MHV integrands
\begin{align}
\Omega_{n,2,2} =\Omega_{n,2,0} \wedge \sum_{i} \Omega\left[\mathfrak{c}_{n,2}^{\{ i \}}\right] \wedge \Omega\left[\Delta^{\{i\}}_{n,2}\right].
\label{eq:fibrations_1}
\end{align}
Since we do not know yet a systematic way of writing the canonical forms for one-loop chambers, this result is still not completely satisfactory. We will remedy this in the next section by rewriting the answer in terms of one-loop leading singularities. 
\subsection{One-Loop Leading Singularities}
In our fibration of fibration approach we made a choice of the ordering of loop variables, and in particular, we consider the second loop geometries parametrised by $y_2$ to be fibered over the first loop parametrised by $y_1$. Knowing the answer for the two-loop integrand \eqref{eq:fibrations_1}, it leads to the natural question of finding all residues of $\Omega_{n,2,2}$ when $y_2$ is completely localised. There are three types of such residues:
\begin{itemize}
\item composite residues $\smash{\Res_{y_2=x_a}\Omega_{n,2,2}}$ when $y_2$ is localised at one of the vertices $x_a$ of the null polygon,
\item residues $\Res_{y_2=q_{aa+1bb+1}^+}\Omega_{n,2,2}$ when $y_2$ is localised at one of the quadruple-cut vertices of the one-loop fiber geometry,
\item residues $\Res_{y_2=q_{aa+1bby_1}^+}\Omega_{n,2,2}$ which corresponds to the remaining vertices of the two-loop fiber geometry.
\end{itemize} 
To expose these residues, it is natural to reorganise \eqref{eq:fibrations_1} and collect terms corresponding to the vertices of the two-loop fiber to arrive at a form for the two-loop integrand given by
\begin{align}
\Omega_{n,2,2} = \sum_{v \in \mathcal{V}^{(2)}_{n,2}} \mathcal{O}_v \wedge \omega_v\,,
\end{align}
where $\mathcal{V}_{n,2}^{(2)}$ contains all vertices that are allowed for the two-loop fiber geometry for given $n$.
We refer to the prefactors $\mathcal{O}_v$ as {\it one-loop leading singularities} corresponding to point $v$. As already demonstrated in \eqref{eq:two_loop_osc_5} for the five-point example, the one-loop leading singularities will generically be given by differences of one-loop amplitudes evaluated for null-polygons with a lower number of points involving both the $x_a$ and $\ell^*_{ab}$. First, using \eqref{eq:two_loop_fiber_form_cp} for the canonical form of the two-loop fibers into \eqref{eq:fibrations_1} we find 
\begin{align}\label{eq:rewriting_two_loops}
\Omega_{n,2,2} = \Omega_{n,2,1} \wedge \sum_{a=1}^n \omega^\Box_{a-1aa+1y_1}+\sum_{(y_1-\ell^*_{ab} )^2>0} \left(\sum_{i\in \mathfrak{C}_{ab}}\Omega\left[\mathfrak{c}_{n,2}^{\{i\}}\right]\right)\wedge\cpp_{aa+1bb+1y_1},
\end{align}
where in the first contribution we used the fact that $\omega_{a-1aa+1y_1}^{\Box}$ is present in \eqref{eq:two_loop_fiber_form_cp} for all chambers, and $\mathfrak{C}_{ab}$ is the set of all chambers for which $\ell_{ab}^*$ is the vertex of the two-loop fiber $\Delta_{n,2}^{\{i\}}$. 
The second term in \eqref{eq:rewriting_two_loops} can be further simplified, leading to a surprisingly simple formula for two-loop integrand for MHV amplitudes
\begin{align}
\frac{\Omega_{n,2,2}}{\Omega_{n,2,0}} =  \Omega_{n,2}^{(1)} \wedge\sum_{a=1}^n \omega^\Box_{a-1aa+1y_1} +\sum_{\substack{1\leq a<b\leq n\\|a-b|>1}}\left( \Omega_{|{\bf x}_{ab}|,2}^{(1)}({\bf x}_{ab})+\Omega_{|{\bf x}_{ba}|,2}^{(1)}({\bf x}_{ba}) \right) \wedge  \cpp_{aa+1bb+1y_1}\,,
\label{eq:two_loop_recurse}
\end{align}
where ${\bf x}_{ab} = \{ x_{a+1}, \ldots,x_b,q^+_{aa+1bb+1} \}$, $ {\bf x}_{ba} = \{ x_{b+1}, \ldots,x_a,q^+_{bb+1aa+1} \}$, $|{\bf x}_{ab}|$ is the cardinality of the set and $\Omega_{n,2}^{(1)}=\Omega_{n,2,1}/\Omega_{n,2,0}$ is the one-loop canonical form with the tree-level contribution removed. The one-loop leading singularities can then be easily extracted from \eqref{eq:two_loop_recurse} by expanding the chiral pentagon contributions.
Note the similarities of this formula to the chiral pentagon expansion \cite{Bourjaily:2013mma} of the one-loop integrand which reads
\begin{align}
\frac{\Omega_{n,2,1}}{\Omega_{n,2,0}} = \sum_{a=1}^n \omega^{\Box}_{a-1aa+1x} + \sum_{\substack{1\leq a<b\leq n\\|a-b|>1}} \cpp_{aa+1bb+1x},
\end{align}
where now the arbitrary point $x$ is taken to be the loop momentum $y_1$ and generically each term is weighted by a sum of one-loop integrands with a lower number of points. 

 \section{Conclusions and Outlook}
\label{sec:conc}
 At one loop the idea of fibrations introduced in \cite{Ferro:2023qdp} relied upon splitting the space of tree-level kinematics into equivalence classes called tree-level chambers. The defining feature of the tree-level chambers was that for any point inside it the resulting one-loop fibers (i.e. the region to which the first loop momentum is constrained) are combinatorially equivalent. This led to a representation of the one-loop integrand in planar $\mathcal{N}=4$ SYM as a fibration over tree-level, in which each term takes a factorised form. In this paper we have extended this construction beyond one loop and introduced the concept of fibrations of fibrations in order to study the two-loop integrand for MHV amplitudes. Extending the original fibration idea, the fibration of fibration asks for the one-loop fiber itself to be split into equivalence classes called one-loop chambers. The significance of the one-loop chambers is that their corresponding two-loop fibers, the region to which the second loop momentum is constrained when the tree-level data and the first loop momentum are fixed, are combinatorially equivalent. This led to the main result of the paper, a new formula for the two-loop integrand \eqref{eq:fibrations} as an iterated fibration over the tree-level kinematics and the position of the first loop momentum. Our result required a full understanding of how the one-loop fiber decomposes into one-loop chambers, and made a connection to the work of \cite{Parisi:2021oql} where similar decompositions of the hypersimplex and $m=2$ amplituhedron were considered. We have also presented simple all-multiplicity formulae for the canonical forms of the two-loop fibers as a sum over box and chiral pentagon integrals, and provided examples of how the canonical forms of the one-loop chambers and two-loop fibers can both be calculated as a sum over their vertices. Finally, by collecting terms coming from the same box or chiral pentagon, we were able to give a new explicit formula \eqref{eq:two_loop_recurse} for the two-loop MHV integrand to all $n$ in terms of one-loop leading singularities.

There are many interesting avenues of future exploration originating from our work. First, it would be interesting to further explore the connection between the one-loop chambers defined here and the results of \cite{Parisi:2021oql}. An interesting question to ask in this direction would be whether the canonical forms of the one-loop chambers can be calculated directly from the, much simpler, canonical differential forms of their corresponding $w$-simplices, for example using the notion of push-forwards of canonical forms \cite{Lukowski:2022fwz}. It would also require one to construct an explicit map from the $w$-simplices to one-loop chambers, which is currently not known. Furthermore, due to the connection between the $m=2$ momentum amplituhedron, a toy model for the momentum amplituhedron $\mathcal{M}_{n,k,0}$, and the hypersimplex \cite{Lukowski:2021amu}, it would further be interesting to explore the connection between the $w$-simplices and one-loop chambers to chambers of the $m=2$ momentum amplituhedron. 

A more pressing question is how to extend our results for the two-loop integrands to higher $k$. Since for $k>2$ the tree-level kinematic space is decomposed into multiple tree-level chambers, then expanding our results beyond MHV amplitudes is exactly the place where we will see the idea of fibration of fibration in full force. It will be particularly interesting to see whether for higher $k$ the canonical forms of the one-loop chambers and the two-loop fiber geometries can still be calculated as a sum over their vertices. 

Another important direction is to explore the fibration of fibration framework at higher loops. We point out that the four-point MHV integrand would already be an interesting but accessible case to study, since there is only one one-loop chamber in this case. Preliminary investigations show that the two-loop fiber decomposes into $68$ {\it two-loop chambers}, suggesting the three-loop integrand can be written in the schematic form
\begin{align}
\Omega_{4,2,3} = \Omega_{4,2,1} \wedge \sum_{i=1}^{68} \Omega[\text{two-loop chamber}_i] \wedge \Omega[\text{three-loop fiber}_i].
\end{align} 
An alternative target for higher loop investigations would be to consider the {\it negative geometries} of \cite{Arkani-Hamed:2021iya,Brown:2023mqi} that allow for some of the mutual positivity conditions between loop momenta to be relaxed. As a starting point, one could consider the mutual positivity conditions encoded by {\it trees in loop space} for all MHV$_n$ in the fibration of fibration formalism. 

Looking beyond $\mathcal{N}=4$ sYM, the null-cone geometries have already been studied in the ABJM theory at one loop \cite{Lukowski:2023nnf}. It is therefore natural to address the problem of finding higher loop geometries in the fibration of fibration framework also there. It might require one to get a better grasp of the two-loop case beyond MHV amplitudes, however the ABJM theory provides another playground, often simplified because of the fact that it is a three-dimensional theory, that is worth exploring.
Moving beyond the realm of scattering amplitudes, another interesting positive geometry where the ideas from this paper could be applied is the correlahedron \cite{Eden:2017fow}, which encodes the correlation functions of stress-energy multiplets in $\mathcal{N}=4$ sYM.  Recently, multi-loop correlahedron canonical forms were studied as fibration over the tree-level kinematics \cite{He:2024xed}. It would be interesting to study the results presented there from the fibrations of fibrations perspective.
 
 \section*{Acknowledgements}
 We would like to thank Jara Trnka, Lauren Williams, Andrew McLeod and Jacob Bourjaily for useful discussions. LF and TL gratefully acknowledge support from the Simons Center for Geometry and Physics, Stony Brook University at which some of the research for this paper was performed. LF, TL and JS gratefully acknowledge support from CMSA Harvard.


\bibliographystyle{nb}

\bibliography{N4_lightcones}

\begin{thebibliography}{10}
\ifx\href\asklfhas\newcommand{\href}[2]{#2}\fi
\ifx\arxivref\asklfhas\newcommand{\arxivref}[2]{\href{http://arxiv.org/abs/#1}{#2}}\fi
\ifx\doiref\asklfhas\newcommand{\doiref}[2]{\href{http://dx.doi.org/#1}{#2}}\fi
\raggedright
\small
\parskip 0pt

\bibitem{Arkani-Hamed:2017tmz}
N.~Arkani-Hamed, Y.~Bai and T.~Lam,
\textit{``{Positive Geometries and Canonical Forms}''},
\textsf{\doiref{10.1007/JHEP11(2017)039}{JHEP~1711,~039~(2017)}},
\texttt{\arxivref{1703.04541}{arxiv:1703.04541}}.

\bibitem{Arkani-Hamed:2013jha}
N.~Arkani-Hamed and J.~Trnka,
\textit{``{The Amplituhedron}''},
\textsf{\doiref{10.1007/JHEP10(2014)030}{JHEP~1410,~030~(2014)}},
\texttt{\arxivref{1312.2007}{arxiv:1312.2007}}.

\bibitem{He:2023rou}
S.~He, Y.-t.~Huang and C.-K.~Kuo,
\textit{``{The ABJM Amplituhedron}''},
\texttt{\arxivref{2306.00951}{arxiv:2306.00951}}.

\bibitem{Arkani-Hamed:2017mur}
N.~Arkani-Hamed, Y.~Bai, S.~He and G.~Yan,
\textit{``{Scattering Forms and the Positive Geometry of Kinematics, Color and
  the Worldsheet}''},
\textsf{\doiref{10.1007/JHEP05(2018)096}{JHEP~1805,~096~(2018)}},
\texttt{\arxivref{1711.09102}{arxiv:1711.09102}}.

\bibitem{Arkani-Hamed:2024nhp}
N.~Arkani-Hamed, Q.~Cao, J.~Dong, C.~Figueiredo and S.~He,
\textit{``{NLSM $\subset$ Tr$(\phi^3)$}''},
\texttt{\arxivref{2401.05483}{arxiv:2401.05483}}.

\bibitem{Arkani-Hamed:2023jry}
N.~Arkani-Hamed, Q.~Cao, J.~Dong, C.~Figueiredo and S.~He,
\textit{``{Scalar-Scaffolded Gluons and the Combinatorial Origins of Yang-Mills
  Theory}''},
\texttt{\arxivref{2401.00041}{arxiv:2401.00041}}.

\bibitem{Damgaard:2019ztj}
D.~Damgaard, L.~Ferro, T.~Lukowski and M.~Parisi,
\textit{``{The Momentum Amplituhedron}''},
\textsf{\doiref{10.1007/JHEP08(2019)042}{JHEP~1908,~042~(2019)}},
\texttt{\arxivref{1905.04216}{arxiv:1905.04216}}.

\bibitem{Ferro:2022abq}
L.~Ferro and T.~Lukowski,
\textit{``{The Loop Momentum Amplituhedron}''},
\textsf{\doiref{10.1007/JHEP05(2023)183}{JHEP~2305,~183~(2023)}},
\texttt{\arxivref{2210.01127}{arxiv:2210.01127}}.

\bibitem{Ferro:2023qdp}
L.~Ferro, R.~Glew, T.~Lukowski and J.~Stalknecht,
\textit{``{Prescriptive unitarity from positive geometries}''},
\textsf{\doiref{10.1007/JHEP03(2024)001}{JHEP~2403,~001~(2024)}},
\texttt{\arxivref{2308.02438}{arxiv:2308.02438}}.

\bibitem{Lukowski:2023nnf}
T.~Lukowski and J.~Stalknecht,
\textit{``{The ABJM Momentum Amplituhedron -- ABJM Scattering Amplitudes From
  Configurations of Points in Minkowski Space}''},
\texttt{\arxivref{2306.07312}{arxiv:2306.07312}}.

\bibitem{Bourjaily:2013mma}
J.~L.~Bourjaily, S.~Caron-Huot and J.~Trnka,
\textit{``{Dual-Conformal Regularization of Infrared Loop Divergences and the
  Chiral Box Expansion}''},
\textsf{\doiref{10.1007/JHEP01(2015)001}{JHEP~1501,~001~(2015)}},
\texttt{\arxivref{1303.4734}{arxiv:1303.4734}}.

\bibitem{Arkani-Hamed:2010zjl}
N.~Arkani-Hamed, J.~L.~Bourjaily, F.~Cachazo, S.~Caron-Huot and J.~Trnka,
\textit{``{The All-Loop Integrand For Scattering Amplitudes in Planar N=4
  SYM}''},
\textsf{\doiref{10.1007/JHEP01(2011)041}{JHEP~1101,~041~(2011)}},
\texttt{\arxivref{1008.2958}{arxiv:1008.2958}}.

\bibitem{Arkani-Hamed:2010pyv}
N.~Arkani-Hamed, J.~L.~Bourjaily, F.~Cachazo and J.~Trnka,
\textit{``{Local Integrals for Planar Scattering Amplitudes}''},
\textsf{\doiref{10.1007/JHEP06(2012)125}{JHEP~1206,~125~(2012)}},
\texttt{\arxivref{1012.6032}{arxiv:1012.6032}}.

\bibitem{Bourjaily:2015jna}
J.~L.~Bourjaily and J.~Trnka,
\textit{``{Local Integrand Representations of All Two-Loop Amplitudes in Planar
  SYM}''},
\textsf{\doiref{10.1007/JHEP08(2015)119}{JHEP~1508,~119~(2015)}},
\texttt{\arxivref{1505.05886}{arxiv:1505.05886}}.

\bibitem{Parisi:2021oql}
M.~Parisi, M.~Sherman-Bennett and L.~Williams,
\textit{``{The m=2 amplituhedron and the hypersimplex: signs, clusters,
  triangulations, Eulerian numbers}''},
\texttt{\arxivref{2104.08254}{arxiv:2104.08254}}.

\bibitem{Arkani-Hamed:2017vfh}
N.~Arkani-Hamed, H.~Thomas and J.~Trnka,
\textit{``{Unwinding the Amplituhedron in Binary}''},
\textsf{\doiref{10.1007/JHEP01(2018)016}{JHEP~1801,~016~(2018)}},
\texttt{\arxivref{1704.05069}{arxiv:1704.05069}}.

\bibitem{Lukowski:2020dpn}
T.~Lukowski, M.~Parisi and L.~K.~Williams,
\textit{``{The Positive Tropical Grassmannian, the Hypersimplex, and the m = 2
  Amplituhedron}''},
\textsf{\doiref{10.1093/imrn/rnad010}{Int.~Math.~Res.~Not.~2023,~16778~(2023)}},
\texttt{\arxivref{2002.06164}{arxiv:2002.06164}}.

\bibitem{Lukowski:2022fwz}
T.~Lukowski, R.~Moerman and J.~Stalknecht,
\textit{``{Pushforwards via scattering equations with applications to positive
  geometries}''},
\textsf{\doiref{10.1007/JHEP10(2022)003}{JHEP~2210,~003~(2022)}},
\texttt{\arxivref{2206.14196}{arxiv:2206.14196}}.

\bibitem{Lukowski:2021amu}
T.~Lukowski and J.~Stalknecht,
\textit{``{The hypersimplex canonical forms and the momentum amplituhedron-like
  logarithmic forms}''},
\textsf{\doiref{10.1088/1751-8121/ac62ba}{J.~Phys.~A~55,~205202~(2022)}},
\texttt{\arxivref{2107.07520}{arxiv:2107.07520}}.

\bibitem{Arkani-Hamed:2021iya}
N.~Arkani-Hamed, J.~Henn and J.~Trnka,
\textit{``{Nonperturbative negative geometries: amplitudes at strong coupling
  and the amplituhedron}''},
\textsf{\doiref{10.1007/JHEP03(2022)108}{JHEP~2203,~108~(2022)}},
\texttt{\arxivref{2112.06956}{arxiv:2112.06956}}.

\bibitem{Brown:2023mqi}
T.~V.~Brown, U.~Oktem, S.~Paranjape and J.~Trnka,
\textit{``{Loops of loops expansion in the amplituhedron}''},
\textsf{\doiref{10.1007/JHEP07(2024)025}{JHEP~2407,~025~(2024)}},
\texttt{\arxivref{2312.17736}{arxiv:2312.17736}}.

\bibitem{Eden:2017fow}
B.~Eden, P.~Heslop and L.~Mason,
\textit{``{The Correlahedron}''},
\textsf{\doiref{10.1007/JHEP09(2017)156}{JHEP~1709,~156~(2017)}},
\texttt{\arxivref{1701.00453}{arxiv:1701.00453}}.

\bibitem{He:2024xed}
S.~He, Y.-t.~Huang and C.-K.~Kuo,
\textit{``{All-Loop Geometry for Four-Point Correlation Function}''},
\texttt{\arxivref{2405.20292}{arxiv:2405.20292}}.

\end{thebibliography}

\end{document}